\documentclass[preprint]{aastex}

\usepackage{amsmath}
\usepackage{graphicx}

\shorttitle{AO O Star Companion Survey}
\shortauthors{Turner et~al.}

\begin{document}

\title{Adaptive Optics Photometry and Astrometry of Binary Stars. III. A Faint
Companion Search of O-Star Systems\altaffilmark{1}}
\author{Nils H. Turner, Theo A. ten~Brummelaar}
\affil{Center for High Angular Resolution Astronomy, Georgia State
University, \\
The CHARA Array, Mount Wilson CA 91023}
\email{nils@chara-array.org, theo@chara-array.org}

\author{Lewis C. Roberts, Jr.}
\affil{The Boeing Company, 535 Lipoa Pkwy, Ste 200 Kihei HI
96753\altaffilmark{2}}
%\email{lewis.c.roberts@boeing.com}
\email{lewis.c.roberts@jpl.nasa.gov}

\author{Brian D. Mason, William I. Hartkopf}
\affil{United States Naval Observatory, Department of Astrometry,
Washington DC}
\email{bdm@usno.navy.mil, wih@usno.navy.mil}

\and

\author{Douglas R. Gies}
\affil{Center for High Angular Resolution Astronomy, Department of 
Physics and Astronomy, \\
Georgia State University, P. O. Box 4106, Atlanta, GA  30302-4106}
\email{gies@chara.gsu.edu}

\altaffiltext{1}{Based on observations made at the Maui Space Surveillance
System operated by Detachment 15 of the US Air Force Research Laboratory's
Directed Energy Directorate.}

\altaffiltext{2}{Current address, Jet Propulsion Laboratory, 4800 Oak Grove
Drive, Pasadena CA 91109}

\addtocounter{footnote}{2}

\section{INTRODUCTION}

Binary star systems are frequently the only source for the fundamental
determination of the most basic property of a star, its mass.
The masses of a sample of stars, when determined with sufficient accuracy,
serve as crucial tests of our theoretical understanding of stellar formation,
structure, and evolution.
By combining information from the angular orbit projected on the sky (or from
a light curve in the case of an eclipsing system) with the spectroscopic orbit,
one can determine the distance and masses of the individual stars.
In the cases of the most massive stars, accurate masses are even more
important, not only because massive stars are relatively rare, but because
these stars (initial masses $>20 M_\odot$) play critical roles in galaxies.
Massive stars serve as signposts of star formation in galaxies
\citep{1995AJ....110.2715M}.
Current theory suggests that high mass stars are born in loosely bound star
clusters which disperse after a few Myr.
These birth conditions suggest that high mass stars might often have
companions, as for example, found around $\gamma^{2}$ Velorum
\citep{2000MNRAS.313L..23P}.
Some of these could be very close and gravitationally bound to the O star,
which is suggested by the prevalence of binary and triple systems among young
O stars \citep{MasonEtAl}.

\cite{MasonEtAl} performed a speckle interferometry survey of Galactic 
O-type stars for close companions, specifically looking for differences in 
the multiplicity frequencies amongst the cluster, field, and runaway O-type 
star populations, as well as the distribution of orbital periods.
They did their survey in the $V$-band using speckle interferometry, which is
sensitive to the detection of a companion if the projected separation is in
the range $0\farcs035 < \rho < 1\farcs5$ and the magnitude difference
$\triangle m_{V} < 3$.
Our survey is in the $I$-band (which will slightly emphasize redder companions)
and extends the dynamic range of their work. 
Due to the availability of only one photometric band for this survey, we are
unable to determine physical associations for the new companions with any
certainty, and we can only address the likelihood that perhaps some of the
companions are gravitationally bound.
The results presented here invite further investigation into the O stars and
their potential to shed light on star formation processes.

\section{OBSERVATIONS}

The data were taken using the adaptive optics (AO) system and Visible Imager
(VisIm) camera of the Advanced Electro-Optical System (AEOS) 3.6-m telescope
at the Maui Space Surveillance System, located on Haleakala.
The data were collected during four separate observing runs between 2001
February and 2002 September.
In addition to the dedicated observing runs, some observations were taken as
part of a queue-scheduled observation program between 2001 May and 2005
November.

\subsection{AEOS Telescope and Adaptive Optics System}

The AEOS telescope is an altitude-azimuth (alt-az) design.
While the telescope is very flexible and supports several different optical 
configurations \citep{RobertsNeyman}, of interest here is the Cassegrain 
configuration which feeds a 726-m focal length, $f/200$ beam (with a 
field of view of $61\farcs9$) through a Nasmyth port coud\'{e} path to a fold
flat below the telescope which directs the beam onto a custom optical table
that supports the main components of the AO system.
For this project, the short-wavelength light (400 -- 700 nm) is sent to the
wavefront sensor (WFS) and tracker camera beam trains, while the 700 nm to
5 $\mu$m light is sent to the VisIm, which is used as the science detector.

The heart of the AO system is a Shack-Hartmann WFS driving a 941-actuator
Xinetics deformable mirror (DM) with a stroke of $\pm4 \mu$m.
Optics inside the WFS produce a pupil image on a Hartmann lens array.
Each generated Hartmann spot is then imaged onto a 4$\times$4 group of pixels
on the WFS CCD.
These pixel values can be used directly or binned (for higher throughput) to
generate the wavefront slopes.
The WFS CCD is a Lincoln Labs frame-transfer device with 16 output ports,
capable of frame rates of 0.2 to 2.5 kHz.
The wavefront slopes are calculated from the Hartmann spot values by means of
many digital signal processors working in parallel, feeding these data to a
system which reconstructs the wavefront, taking into account alt-az induced
image rotation and WFS-to-DM-actuator mapping artifacts.

The VisIm camera is described in detail in \cite{RobertsNeyman}.
In short, it operates from 700 to 1050 nm; is equipped with an atmospheric
dispersion corrector; has a two-mode image derotator (zenith at a fixed
position in the image or celestial north at a fixed position in the image);
and, for this project, has a $10''$ field of view ($0\farcs020$ pixel$^{-1}$).
The detector is a 512$\times$512 pixel EEV CCD with a dark current of
22 $e^{-}$ pixel$^{-1}$ s$^{-1}$ (at $-$40$^{\circ}$ C) and a read noise of
12 $e^{-}$ rms.
The camera output is digitized to 12 bits with 10 $e^{-}$ per digital number.

\subsection{Object Selection}

The object list started out as all the components in the 228 systems listed 
in \cite{MasonEtAl}.
They were then culled for the effective magnitude limit of the AEOS AO system
(about $m_{V} = 8$), and for the declination limit of acceptable AO
correction (objects that at some time during the year get above $30^{\circ}$
elevation at Haleakala, i.e.,  a declination greater than about $-45^{\circ}$).
This reduced the list to 164 objects.
We added seven more targets from the Galactic O Star
Catalog\footnote{{\tt http://www-int.stsci.edu/$\sim$jmaiz/research/GOS/GOSmain.html}}
\citep{GOS} that are within our adopted magnitude and declination limits.
We actually observed, at least once, 116 of the 171 objects in our list.
These observations are fairly consistently distributed among the three O-star
populations, sampling 63\% of the cluster membership stars, 78\% of the field
stars, and 88\% of the runaway stars in our list of potential targets.

%Numbers:
%Cluster -- 81 observed of 128
%Field   -- 21 observed of 27
%Runaway -- 14 observed of 16

\subsection{Data Collection and Reduction}

\begin{figure}
\plotone{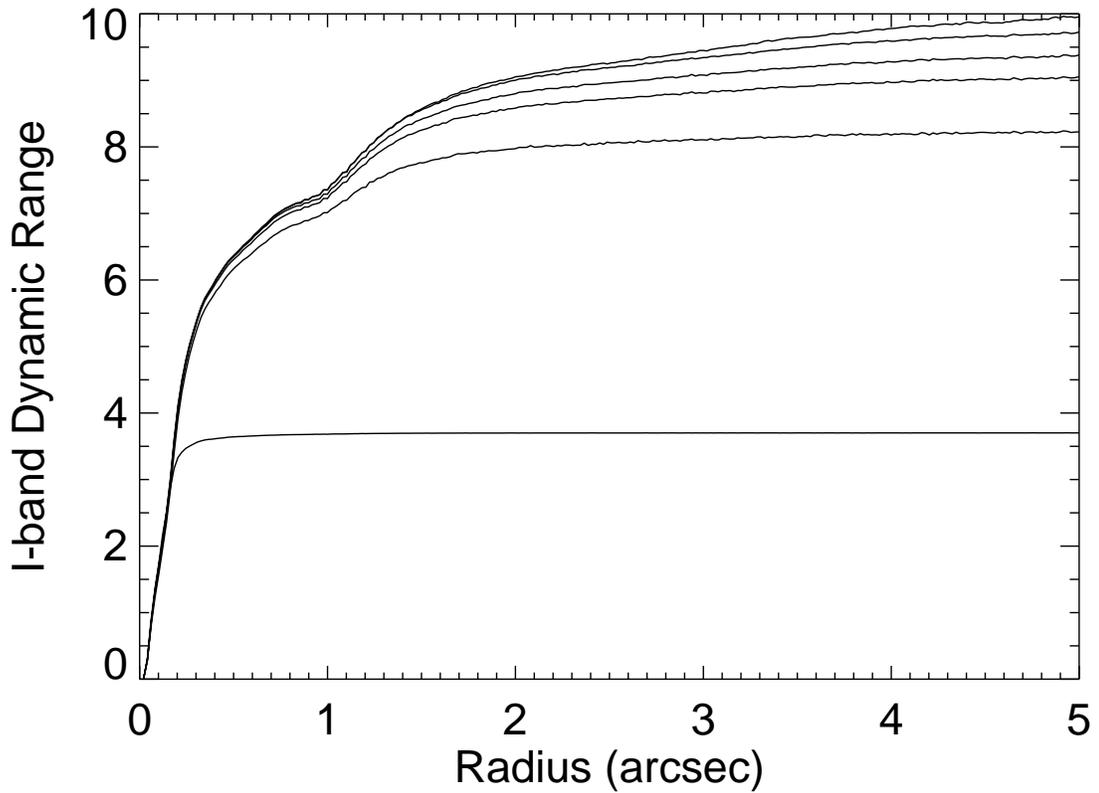}
\caption[Detection Dynamic Range vs. Angular Distance]{Assuming that a
Gaussian-shaped peak, $3\sigma$ detection above background constitutes a
positive detection of a companion, this plot shows the rough dynamic range 
of the AEOS AO system as a function of radius for different numbers of 
summed images. From top to bottom, the curves represent the results of 1000,
250, 100, 50, 10, and 1 summed frame(s). The ``shelf'', most apparent in the
upper curves from about 0.7 to 1 arcsec in radius, is due to AO system 
artifacts.}
\label{fig:dynamic}
\end{figure}

Since VisIm is only a 12-bit camera, a stellar image can overflow the 
digitizer in a rather short exposure time.
If the frame saturates, then it is omitted from the final image.
Frequently, this saturated frame will represent a period of particularly good
seeing.
In order to keep as many of these ``good seeing'' frames as possible, we set
the exposure time such that the average peak value was about 75\% of the
full-well depth.
We then  built up the signal-to-noise ratio by taking many frames, then
weighting and summing them.
Figure~\ref{fig:dynamic} shows the effect of this summing on the detectable
magnitude difference for 1, 10, 50, 100, 250, and 1000 frame(s).

On the basis of this plot, we aimed to take at least 1000 frames of each
object, although time constraints, weather, and object brightness occasionally
limited us to a lower frame count.

The VisIm is a frame transfer camera and is therefore unable to take a bias
frame.
To compensate, a bias frame was created by taking many dark exposures at a
variety of exposure times ranging from 10 ms to several minutes.
A linear fit was made to the signal from each pixel as a function of exposure
time; the $y$-intercepts became the bias frame while the slopes became the dark
frame.
Prior to 2005, flat-field frames were created using the twilight sky to 
evenly illuminate the VisIm detector.
\cite{amos2001} found that sky flat-field frames varied at about the 1\% level,
so after 2005, flat-field frames were generated using an internal calibration
sphere; this lowered the frame-to-frame variation below the 0.01\% level.
From this point, all data frames were debiased, dark-subtracted and
flat-fielded in the conventional way.

For a given sequence of data frames, a weighted shift-and-add algorithm was
used to create the final image.
The weighted shift-and-add algorithm \citep{spie3353-391} is a modification of
the traditional image-stacking algorithm that takes seeing conditions in each
individual frame into account.
Frames with higher peak values (which represent better seeing) influence the
final image (by means of weighting) more than frames with lower peak values.

The fitting algorithm is described in detail in \cite{diffmag1,diffmag2}.
In short, the point-spread function (PSF) used to represent the system
performance is that of the primary star in the image, with a few modifications;
near the primary, the PSF is a pixel-for-pixel table of numbers, while farther
out the PSF is a radially-symmetric series of values.
This PSF is fitted to the primary and secondary components and iterated until
the intensity ratio converges.

\subsection{Detection Sensitivity}

Figure~\ref{fig:dynamic} only gives a qualitative sense of the companion
detection sensitivity of the system.
Cross comparisons of the reduced images, especially of the same object taken on
different nights, show different detection limits.
This is primarily due to variations in AO performance which is primarily
attributed to variations in the atmospheric turbulence profile.
In order to quantify the sensitivity of the reduced images, we have created a
variation of the ``dynamic range map'' technique described in
\cite{2007ApJ...654..633H} which defines the dynamic range of a given position
in a 2-D image as the faintest companion detectable at that position to the
$5\sigma$ level.
In our version, we construct a map the same size as each reduced image.
The intensity level of each pixel in the map is set to five times the RMS
intensity variation across a patch centered on the corresponding pixel
in the original image.
The patch is a square with lengths equal to the FWHM of the original image.
This produces the dynamic range map in intensity terms, which are then
converted to magnitudes.

\begin{figure}
\plotone{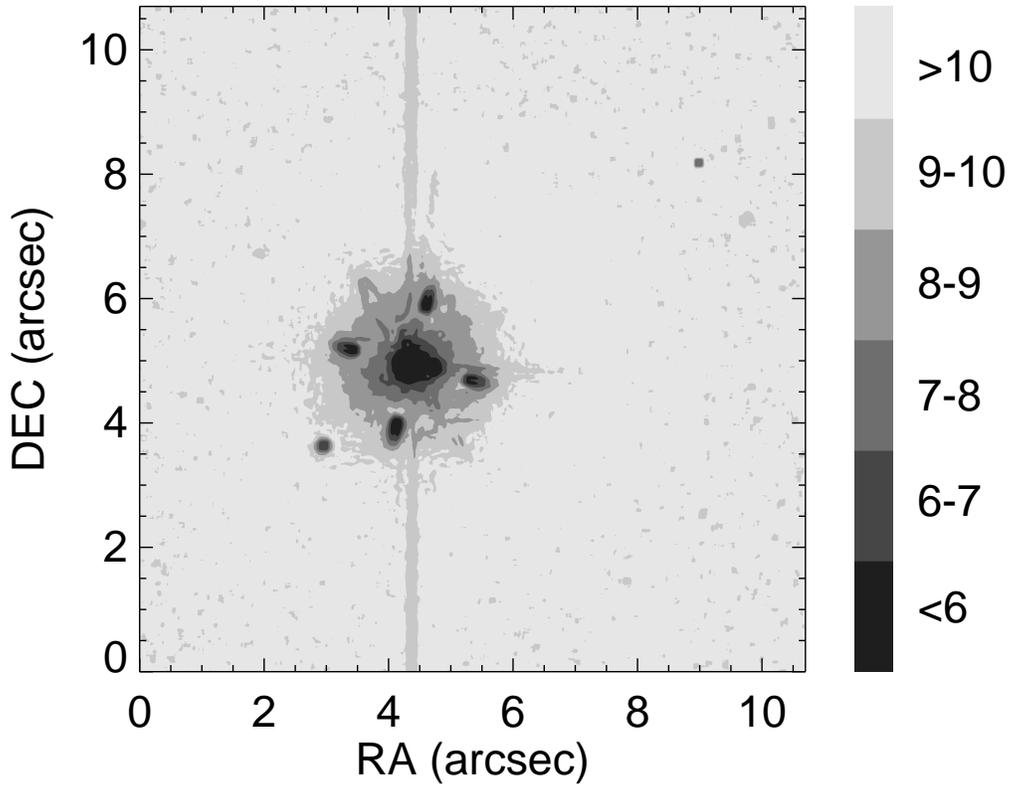}
\caption[2-D Dynamic Range Map]{A sample dynamic range map (in this case,
HD~34656, taken Bess. Year 2003.7360).  The gradient map on the right shows
the $5\sigma$ detection limit for a given pixel of the image in magnitudes.
There are several artifacts in this image which lower the dynamic range. The
largest is the vertical line, which is an artifact of the shutter-less frame
transfer process. There are also four waffle spots caused by the Fried
geometry of the wavefront sensor. There are also hints of the diffraction
pattern caused by the secondary mirror support spiders.  These all increase
the difficulty of finding any companions that coincide with these artifacts.
}
\label{fig:dyn_rang_maps}
\end{figure}

Figure~\ref{fig:dyn_rang_maps} shows an example of this technique applied
to one of the images generated in this survey.

It is apparent from the sample figure that the detection threshold for a
companion is highly spatially variant.
The dynamic range increases with increasing radius from the central star.
There are several artifacts in this image which lower the dynamic range.
The largest is the vertical line, which is an artifact of the shutter-less
frame transfer process.
There are also four waffle spots caused by the Fried geometry of the wavefront
sensor \citep{2005PASP..117..831M}.
There are also hints of the diffraction pattern caused by the secondary mirror
support spiders.
As the AO system performance decreases, the FWHM of the PSF will increase,
which increases the area over which the central PSF causes confusion.
In addition, it also widens the companion's PSF, lowering the contrast between
the two and decreasing the detection rate.
After 2003, the waffle spots were greatly reduced when new reconstructors were
put in place to filter out the unsensed waffle modes.

These maps have several purposes.
For those objects with multiple observations where companions are seen in some
but not all of the data, changes in the dynamic range can help explain the
reasons for these discrepancies.
They also illustrate the limitations of the data, showing where it is almost
impossible to find faint companions.
This is often useful when trying to detect known long-period spectroscopic
companions.

\section{RESULTS}

\subsection{New Pairs and New Measures}
\label{ssec:pairs-measures}

Table~\ref{tbl:merge} lists the 40 new and 24 known pairs detected in the
survey.
Four of the new companions (HD 17505, HD 37468, HD 193793, and HD 217086) are
in systems with previously known components in the AEOS field of view.
The remainder are new companions in systems not previously known to be
multiple (at least in the AEOS field of view).
The table has nine columns which summarize the information gleaned from the
measurements: Column (1) lists the Washington Double Star (WDS) identifier,
columns (2) and (3) identify the object by HD and Hipparcos number, and
column (4) gives the discoverer designation (TRN for new pairs).
Circumstances of the observation are given in columns 5--8.
Column (5) gives the Besselian epoch of the start of the observation, columns
(6) and (7) give the position angle (in degrees, where north is 0$^{\circ}$
and east is 90$^{\circ}$) and separation (in seconds of arc) between the
primary (the survey star -- the brightest star in the image) and the detected
component.
Notice that for many of the objects, the position angle is missing or
ambiguous.
The VisIm derotator is usually in one of two modes, one which keeps zenith up
in the image, and one which keeps north up in the image.
In the observations with more than one or a missing position angle, the state
of the derotator was unknown, leaving the orientation of the image on the sky
unknown.
However, in some cases, position of a known pair resolves the ambiguity while
for the rest, new observations will be required.
Column (8) gives the magnitude difference at $I$-band between the primary and
the detected component.
Finally, column (9) lists the spectroscopic status.
Given the separations detected here, none of the short period spectroscopic
pairs are expected to be detected.
However, listing the spectroscopic binaries helps to form the complete
multiplicity picture.
The indicator in this table is {\bf V} (velocity variable) if one or more close
spectroscopic companions is indicated or {\bf C} (velocity constant) if radial
velocity work indicates no close companion.
The spectroscopic reference is generally given in \cite{MasonEtAl}.

The differential magnitude errors in column~8 of Table~\ref{tbl:merge} were
generated from simulated binary data that we analyzed using the same fitting
algorithm applied to the measured binary data.
To generate the simulated binaries, we started by collecting all the single
star images from this survey (see Table~\ref{tbl:sing}) and those of a
duplicity survey B stars carried out contemporaneously
\citep[see][Table~2]{2007AJ....133..545R}.
We omitted single star images where there were fewer than 250 frames
used to make the final image.
This exercise gave us 167 unique images with PSFs ranging from $0\farcs083$ to
$0\farcs445$, with a decided emphasis on values less than $0\farcs2$.
For each entry in Table~\ref{tbl:merge}, 167 simulated binaries were
constructed using the magnitude difference in column~8 and the pixel positions
of the primary and secondary components in the main image.
We then ran the fitting algorithm on each of these 167 simulated images for
each table entry.
We calculated the magnitude variance in two passes, the first to filter out
outliers on the basis of the variance of the fitted pixel separation of the
simulated binary (namely, any deviation greater than $1\sigma$), and the second
to use this reduced list of simulated fits to generate the final magnitude
variance.
In general, the number of outliers per table entry was less than 10.
The errors quoted in column~8 are the standard errors of these variances.
Errors are not quoted for the separation (column~7), but in all cases, the
standard error is less than $0\farcs04$, usually less than $0\farcs005$.
For errors in position angle (column~6), we adopt the values from
\cite{2007AJ....133..545R}, $\pm2^{\circ}$ for separations less than $1''$,
and $\pm1^{\circ}$ for separations greater than $1''$.
Two effects lead to this uncertainty, the clocking calibration of the dove
prism in the derotator, and the slight natural image rotation during the frame
collection sequence when the derotator is in the fixed zenith position mode.

\subsubsection{Notes to Pairs}

Here we discuss the various binary systems in Table~\ref{tbl:merge} by
WDS identifier and discovery designation.
In the cases of the stars classified as runaways, it is unlikely that the
detected companion is physical (see \S~\ref{sec:discussion}).

\noindent {\bf 02229$+$4129 = TRN\phn\phn10}:
This is a runaway star with a close, low-luminosity companion
\citep{2005ApJ...621..978B}.

\noindent {\bf 02407+6117 = TRN\phn\phn12AD}:
The A component is a spectroscopic triple \citep{2003ApJ...595.1124M}.

\noindent {\bf 02511$+$6025 = TRN\phn\phn13AH}:
The A component of HD~17505 is a spectroscopic triple
\citep{2006ApJ...639.1069H}.
It is uncertain if the new component (AH, $\rho \sim$ 4\farcs6) and the known
pair of Table~\ref{tbl:merge} (STF\phn306AB) are both gravitationally bound.
If STF\phn306AB is indeed gravitationally bound, \cite{2006ApJ...639.1069H}
estimate it to have a period of 27,000 years.

\noindent {\bf 03556$+$5238 = HDS\phn494}:
The $\triangle$m was probably a little too large for \cite{MasonEtAl} to detect
this pair, but it was detected in later speckle observations by
\cite{2001AJ....121.3224M}.

\noindent {\bf 03590$+$3548 = TRN\phn\phn16}:
Classified as a runaway star in \cite{1986ApJS...61..419G}.

\noindent {\bf 05163$+$3419 = TRN\phn\phn17Aa}:
Classified as a runaway star in \cite{1986ApJS...61..419G}.  SEI\phn136AB
($\rho \sim$ 8\farcs8, $\triangle$m $\sim$ 3.3, just outside the field of view)
is a likely optical pair, with the B component not physically associated with
this close double.

\noindent {\bf 05207$+$3726 = TRN\phn\phn18}:
One new close pair and three wide pairs are found here. The close one is
designated Aa while the wider pairs are designated AB, AC, and AD. The known
pair SEI\phn201 ($\rho \sim$ 24\farcs7, $\triangle$m $\sim$ 5.9) is outside
the field of view and has been designated AE.

\noindent {\bf 05297$+$3523 = HU\phn\phn217}:
This known, relatively close, pair has been measured since 1900.
The position angle has only changed by $4^{\circ}$ over this time span.

\noindent {\bf 05353$-$0523 = STF\phn748Ca,F}:
The primary is $\theta^1$~Ori~C, the most massive star in the Orion Trapezium.
It has a closer companion discovered by speckle methods
\citep{2007AaA...466..649K}.

\noindent {\bf 05354$-$0525 = CHR\phn249Aa}:
HIPPARCOS \citep{1997hity.book.....P} measured a $\triangle H_{p}$ of
3.23.
\cite{2000AaA...356..141F} found $\triangle B_{T} = 4.37 \pm 0.02$
and $\triangle V_{T} = 3.21 \pm 0.01$.

\noindent {\bf 05387$-$0236 = TRN\phn\phn19AF}:
The new component (AF) is at a separation nicely nestled between BU\phn1032AB
and STF\phn762AC, at a value which would be consistent with a hierarchical
arrangement.
The new TRN~19AF component is much brighter in the mid-infrared where it was
first discovered \citep{2003AaA...405L..33V} and it is probably a low mass
K-star surrounded by a circumstellar disk \citep{2007AaA...466..917C}.
See \S~\ref{orbit_discussion} below for a discussion of the orbit of BU~1032AB.

\noindent {\bf 05407$-$0157 = STF\phn774Aa-B}:
This pair has an orbit \citep{1967MiWie..13...49H} with residuals of:
O$-$C = $-5^{\circ}$ and $-$0\farcs03 in position angle and separation.
However, the orbit is classified as ``indeterminate'' (grade 5) in the
{\it 6th Catalog of Orbits of Visual Binary
Stars}\footnote{{\tt http://ad.usno.navy.mil/wds/orb6.html}}, so
these residuals are not indicative of measurement quality.
While the total number of measures has increased by about 50\% since this orbit
was published, the initial elements are of such low quality that even at this
point any sort of improvement above ``indeterminate'' subjective quality is not
yet possible.
Some of the scatter in the residuals may be due to the presence of a closer
companion to $\zeta$~Ori~A found by interferometry \citep{2000ApJ...540L..91H}.

\noindent {\bf 06319$+$0457 = GAN\phn\phn\phn3AB}:
The measure agrees with earlier published data quite well and indicates that
the measure made in 1928 \citep{1931AN....241...33S} is probably erroneous.

\noindent {\bf 06410$+$0954 = STF\phn950Aa-B}:
This pair has only changed by $\sim$9$^{\circ}$ and 0\farcs2 since it was first
resolved by Struve in 1825 \citep{1837AN.....14..249S}.

\noindent {\bf 16550$-$4109 = I\phn\phn\phn576}:
I\phm{888}576 has not been measured since 1934 \citep{Wallenquist1934}, so
identification of this pair with that of Innes is uncertain.

\noindent {\bf 17065$-$3527 = B\phn\phn\phn894}:
This binary, unconfirmed since its first resolution by \cite{vandenBos1928},
has closed slightly.

\noindent {\bf 17158$-$3344 = SEE\phn322}:
This pair has shown noticeable Keplerian motion, completing about 70$^{\circ}$
of orbital revolution since discovery \citep{1898AJ.....18..181S}.
See \S~\ref{orbit_discussion} below for a discussion of the orbit.

\noindent {\bf 17347$-$3235 = HDS2480Ab}:
The pairs, ISO\phn\phn\phn5 and HJ\phn4962 should have been seen as well.
ISO\phn\phn\phn5 was probably missed due to the corrected FWHM.
HJ\phn4962 was probably just out of the field.  

\noindent {\bf 18024$-$2302 = TRN\phn\phn26AH}:
While this pair may be physical, the known double H\phm{8}N\phm{88}40 almost
certainly is not.

\noindent {\bf 18026$-$2415 = RST3149AB}:
This pair has been seen only twice since 1935.
The occultation pair of \cite{1975AJ.....80..689A} was not seen.

\noindent {\bf 18061$-$2412 = B\phn\phn\phn376}:
Unconfirmed since its discovery in 1927 \citep{vandenBos1928}.
The A component is an eclipsing binary with a 4.6 d period
\citep{2007OEJV...72....1O}.

\noindent {\bf 20035$+$3601 = STF2624Aa-B}:
The close pair, MCA\phn\phn59Aa, which is notoriously difficult to detect, was
not seen here.

\noindent {\bf 20181$+$4044 = STF2666Aa-B}:
The close speckle pair CHR\phn\phn96Aa was not recovered.

\noindent {\bf 20205$+$4351 = TRN\phn\phn29AC}:
It is unclear whether there are two new components or a nearby optical pair
with a large relative proper motion detected at two different positions.

\noindent {\bf 21390$+$5729 = BU\phn1143AB}:
The close Aa pair (MIU\phn\phn\phn2) was not recovered, although it may have
been too close.

\noindent {\bf 21449$+$6228 = TRN\phn\phn33AC}:
The visual pair, MLR\phn\phn16, would have been much too wide (17$''$\llap.6)
to detect here.

\noindent {\bf 22393$+$3903 = TRN\phn\phn36AC}:
The visual binary, S\phn\phn\phn813, at $\sim$ 1 arcminute of separation is
much too wide to detect here.

\noindent {\bf 22568$+$6244 = TRN\phn\phn37AC}:
It is hard to imagine a geometry where both this pair and MLR\phn266 would be
dynamically stable.  More than likely one (or both) are optical.

\subsubsection{Comparison of $V$- and $I$-band Magnitude Differences}

\begin{figure}
\centering
\plotone{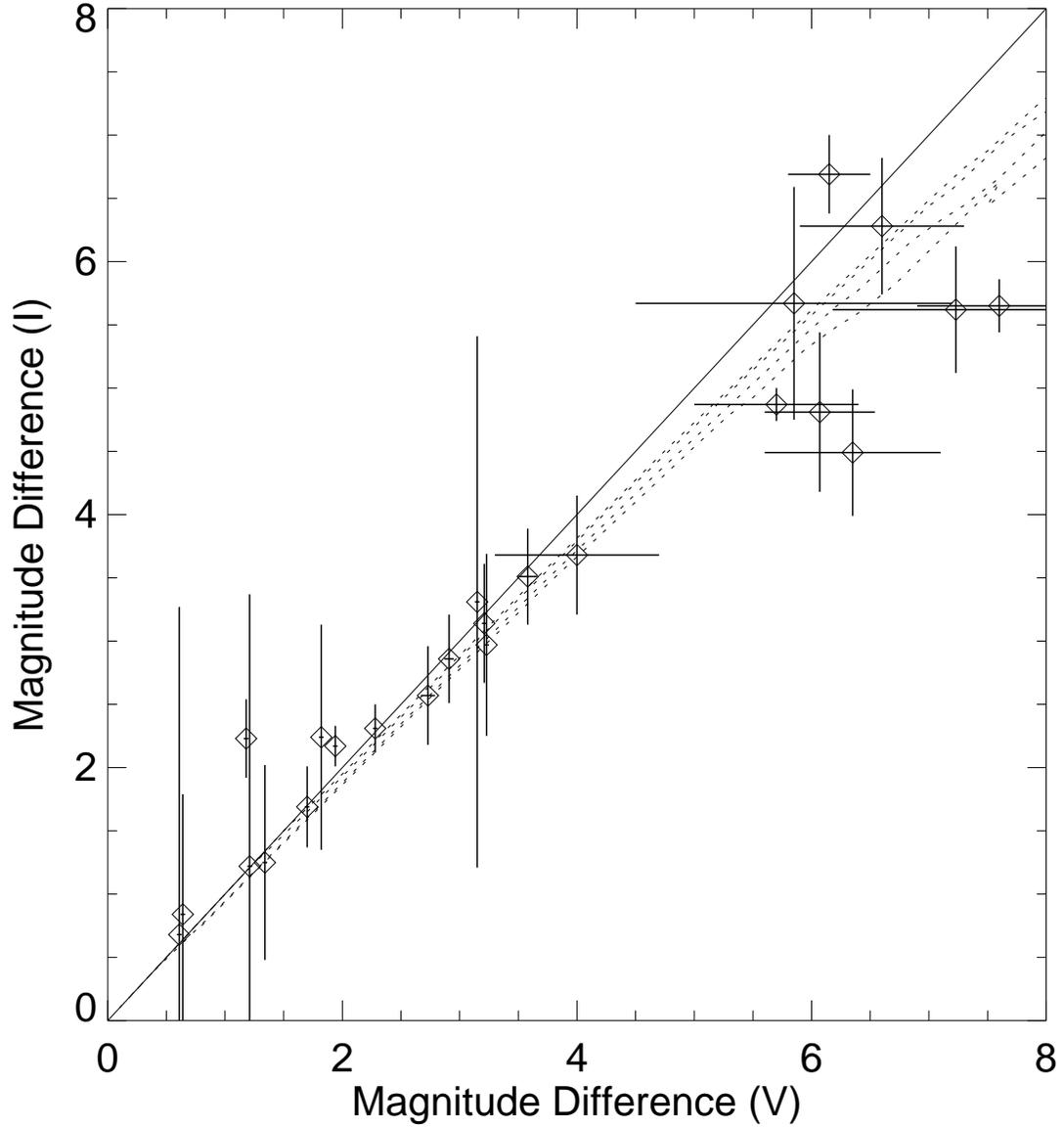}
\caption[$V-I$ Color-Color Diagram]{$V$-band magnitude differences from the WDS
versus $I$-band values from AO.  The solid line indicates a 1:1 relation
between $\triangle V$ and $\triangle I$, while the various dotted curves
illustrate the expected trend assuming main sequence primaries of types O3, O5,
O8, and B0 (from top to bottom, respectively) and a range of later
spectral-type secondaries.}
\label{fig:dvdi}
\end{figure}

Among the O star primaries listed in the WDS, we may estimate the companion
spectral type from the magnitude difference $\triangle V$ for those with small
magnitude differences.
As the magnitude difference gets larger, the uncertainty in the companion
spectral type increases.
However, the situation improves with the availability of magnitude differences
for two filters.
We compare the $V$-band magnitude differences from the WDS with the $I$-band
values we determined by AO.
The result is shown in Figure~\ref{fig:dvdi}.

Error bars for the visual data were based on typical Tycho-2 photometric errors
when available (smaller-$\triangle$m pairs), and on scatter among individual
visual estimates for those (larger-$\triangle$m) systems with more than one
such estimate.
A rough average value of this scatter (0.7 mag) was used for systems with only
one visual $\triangle m_{V}$ estimate.
Mean $\triangle m_I$ values and error bars are given for systems with
multiple observations.
For systems with no $I$-band error estimate a default value of 0.5 magnitudes
was adopted.

This figure shows that $\triangle V \approx \triangle I$ for small magnitude
differences, but $\triangle V$ increases more rapidly at larger values of
$\triangle I$.
This is as expected for our set of O-star targets: the components of
small-$\triangle$m systems are of similar spectral type, but the secondaries
become later in spectral type --- hence redder --- as $\triangle$m increases.

To determine the expected extent of this trend, absolute $V$ and $I$ magnitudes
for main sequence spectral types were extracted from
{\it Allen's Astrophysical Quantities} \citep{APQ2001}.
Values of $\triangle V$ and $\triangle I$ were then determined, assuming
primaries of spectral types O3, O5, O8, and B0 and secondaries covering a wide
range of spectral types.
The results, plotted as dotted lines in Figure~\ref{fig:dvdi}, appear to be in
good agreement with the trend seen in these AO observations and are consistent
with the assumption that the companions are physical.

\subsubsection{Orbits of BU~1032 ($\sigma$~Ori~AB) and SEE~322 (HD~155889)}
\label{orbit_discussion}

\begin{figure}
\plotone{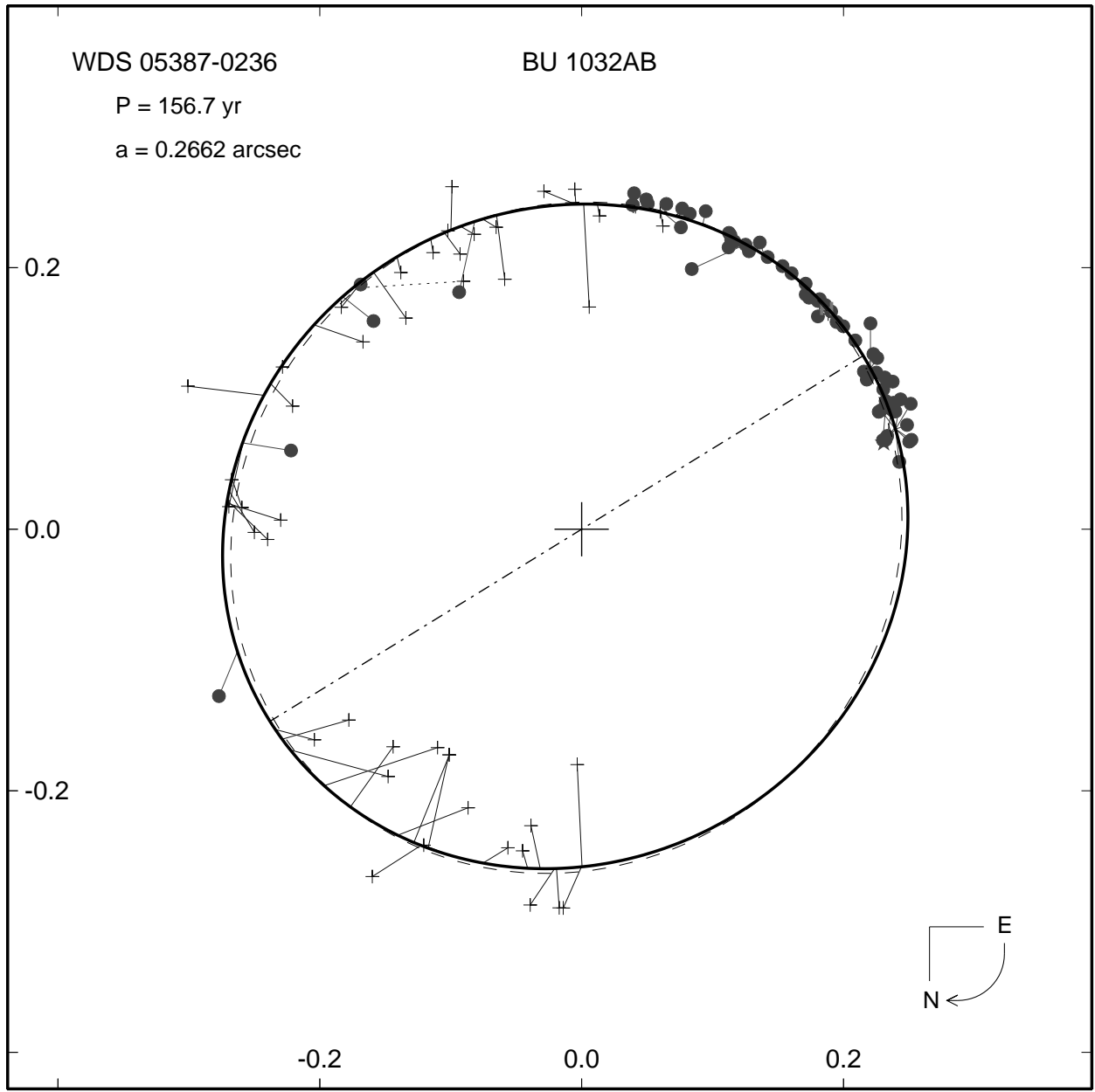}
\caption[Orbit of BU 1032]{Relative visual orbit of BU 1032; the $x$ and $y$
scales are in arcseconds.  The solid curve represents the newly determined
orbit (Table~\ref{tbl:orb_el}).  The dot-dash line indicates the line of nodes.
High-resolution measurements are shown as filled circles (speckle) or filled
stars (AO).  Visual measurements are denoted with plus signs.  The orientation
and direction of motion are indicated in the lower right corner of the plot.
The orbit of \cite{1996AJ....111..370H} is shown as a dashed curve, which
matches the new orbit quite well.}
\label{fig:wds05387-0236}
\end{figure}

\begin{figure}
\plotone{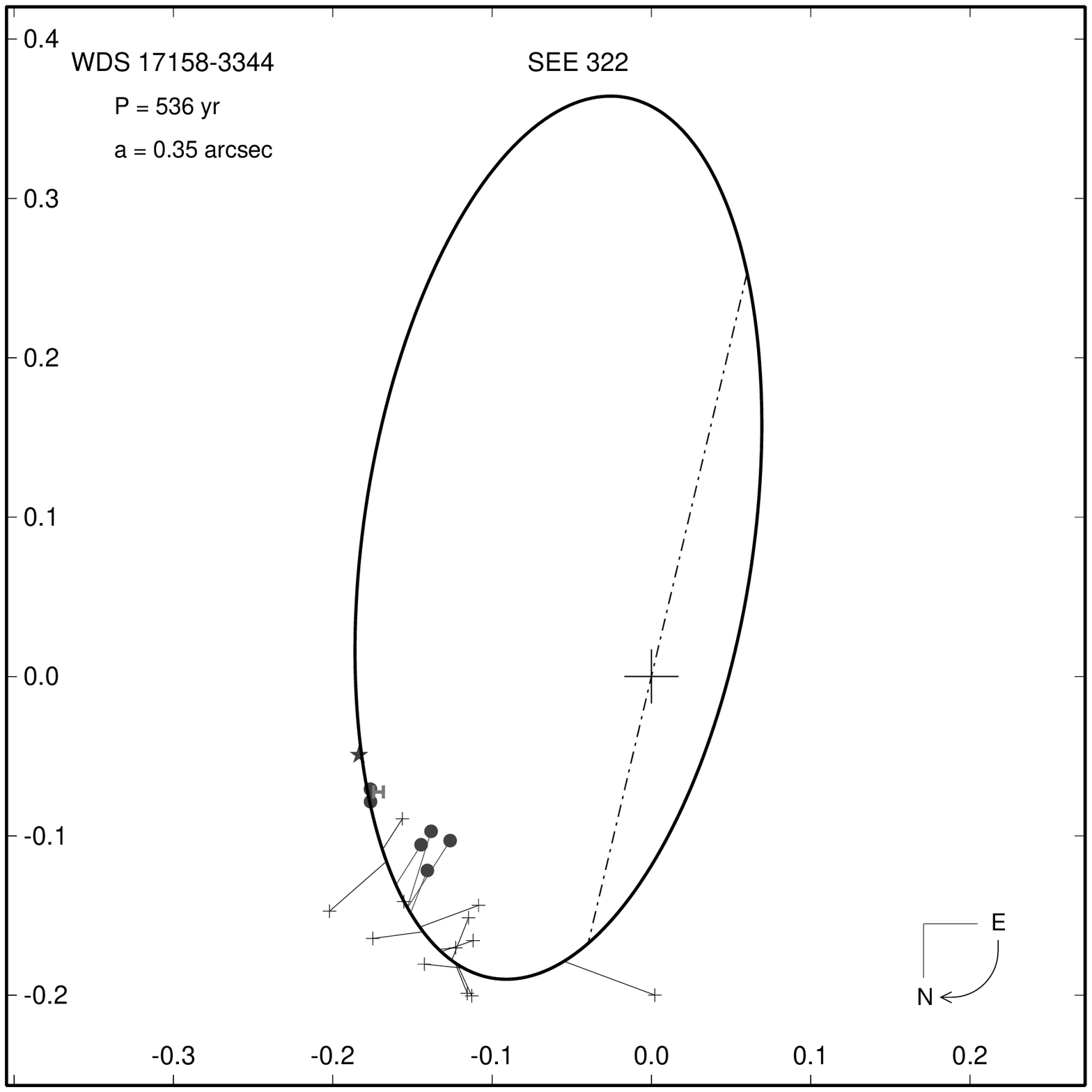}
\caption[Orbit of SEE 322]{Relative visual orbit for SEE~322 in the same format
as Fig.~\ref{fig:wds05387-0236}.  The {\bf H} indicates the measure of
Hipparcos.}
\label{fig:wds17158-3344}
\end{figure}

\cite{1990ebua.conf..419W} provides guidelines as to when a collection of
binary star measures merits the publication of a new orbit.
Following these criteria, we determined that two systems, BU~1032 and SEE~322,
deserved new orbital solutions.
All available measures for these pairs were extracted from the WDS
\citep{WDS2001} and individual measures were weighted following the precepts of
\cite{2001AJ....122.3472H}.
The orbital elements were determined with an iterative three-dimensional
grid-search algorithm \citep{2002AJ....123.1023S}.
For these two pairs, orbital elements are provided in Table~\ref{tbl:orb_el}
and future ephemerides in Table~\ref{tbl:ephem}.
Elements from the \cite{1996AJ....111..370H} orbit for BU~1032 are provided for
comparison.
Due to the preliminary nature of the SEE 322 orbit, errors are not quoted.
These orbits are illustrated in Figures~\ref{fig:wds05387-0236} and
\ref{fig:wds17158-3344}.

Since the publication of the orbital elements of BU\phn1032 by
\cite{1996AJ....111..370H}, the number of measures has increased by about 35\%,
and while the prior orbit adequately fits these data, a new calculation at this
time makes a significant improvement in the orbit quality, as illustrated by
the decrease in formal errors for each of the elements.
The distance and mass of the $\sigma$~Ori~AB system are discussed in a recent
paper by \cite{2008MNRAS.383..750C}.

There is no previous determination of orbital elements for SEE~322.
While the errors and residuals are large, the orbital solution looks promising,
although it is still preliminary since only a limited portion of the orbit is
covered by existing data.
The total mass associated with this solution is very large ($260~M_\odot$ for a
distance of 1.2~kpc; \citealt{MasonEtAl}), which may indicate that revisions are
required in the assumed elements and distance, or that the system contains more
than two stars.

\subsection{Single Star Detections}

In addition to discovering new companions and measuring some parameters of 
previously known stars, we have a number of stars determined to be single under
the specific observing conditions, AO system performance, and field of view.
Table~\ref{tbl:sing} lists these stars.
In this table, the respective columns give the HD number, Hipparcos number,
epoch of observation, full-width at half-maximum of the corrected stellar PSF,
and the V/C code for the spectroscopic binary status
(see \S~\ref{ssec:pairs-measures}).
A larger FWHM is indicative of poorer seeing, a dimmer primary, non-optimal AO
system tuning, or some combination of these factors.

\subsubsection{Notes to Unresolved Systems}
\label{unresolved_notes}

\noindent {\bf HD \phn15137}:
This runaway star is a probable, single-lined spectroscopic binary
\citep{2005ApJ...621..978B}.

\noindent {\bf HD \phn25638}:
The pair ES\phn2603 ($\rho \sim$ 6\farcs5, $\triangle m \sim$ 6.0) was
not detected.
It was probably just out of the field of view.

\noindent {\bf HD \phn30614}:
Classified as a runaway star in \cite{1986ApJS...61..419G}.

\noindent {\bf HD \phn36879}:
This star is a runaway object according to its proper motion
\citep{2005AaA...431L...1M}.

\noindent {\bf HD \phn37042}:
The known components to this star are all far too wide for detection here.
The star $\theta^2$~Ori~B is radial velocity variable according
to \cite{1991ApJS...75..965M}.

\noindent {\bf HD \phn37043}:
Iota Ori is a close speckle pair of great interest as, despite its very close
separation, it was deemed optical in \cite{2004MNRAS.350..615G}.
The speckle pair was last measured very close to the FWHM value, so it is
possible that the component was just barely too close to be detected.

\noindent {\bf HD \phn37366}:
This star is a double-lined spectroscopic binary \citep{2007ApJ...664.1121B}.

\noindent {\bf HD \phn39680}:
Classified as a runaway star in \cite{1986ApJS...61..419G}.
The known pair, S\phm{888}502, is much too wide (46\farcs1) for detection here.

\noindent {\bf HD \phn41161}:
The known pair, ES\phn1234AB, was too wide (10\farcs3) for detection here.
Whether this pair is bound or not is not yet certain.

\noindent {\bf HD \phn45314}:
CHR\phn251 was last measured closer (54 milliarcsec) than the FWHM of the PSF
of our observation.
The A component is an Oe star and a probable velocity variable
\citep{2007PASP..119..742B}.

\noindent {\bf HD \phn52266}:
This star is a probable single-lined spectroscopic binary
\citep{2007ApJ...655..473M}.

\noindent {\bf HD \phn54662}:
This star is a double-lined spectroscopic binary \citep{2007ApJ...664.1121B}.

\noindent {\bf HD \phn60848}:
This Oe star shows emission line variations but no evidence of significant
velocity variability \citep{2007PASP..119..742B}.

\noindent {\bf HD \phn69648}:
Not included in \cite{MasonEtAl} survey.

\noindent {\bf HD    163892}:
These data were taken using an 800-900\,nm filter.

\noindent {\bf HD    164794}:
This star is a probable long-period spectroscopic binary
\citep{2002AaA...394..993R}.

\noindent {\bf HD    167771}:
The known companion, RST3170, last measured in 1940 at 8\farcs3
\citep{1955POMic..11....1R}, would have been outside the field of view. 

\noindent {\bf HD    175876}:
The known, probably optical, pair, HO\phm{88}271, would have been too wide to
detect.

\noindent {\bf HD    191978}:
Not included in \cite{MasonEtAl} survey.
The star shows no evidence of velocity variability \citep{1972AJ.....77..138A}.

\noindent {\bf HD    193443}:
The close pair, A\phm{88}1425AB ($\rho\sim$ 0\farcs1) may have been too close
to resolve, while the wider AC pair ($\rho\sim$ 9\farcs1) may have been outside
the field of view.

\noindent {\bf HD    195592}:
This is a probable single-lined spectroscopic binary
\citep{2007ApJ...655..473M}.

\noindent {\bf HD    216898}:
Not included in \cite{MasonEtAl} survey.

\subsection{2MASS Data Mining Confirmations}

Searches were made for 2MASS \citep{2MASS} companions to all stars on our
observing list, using Vizier and  Aladin, as well as the `data mining'
technique of \cite{2006AJ....132...50W}.
Results are given in Table~\ref{tbl:2mass} and discussed below.

It should first be noted that the separation/$\triangle$m regimes covered by
these two techniques have rather limited overlap.
The ``plate scale'' of the AO detector nominally limits us to companions within
$\sim5''$ of the primary, although this outer limit may extend to as wide as
7\farcs5, depending on orientation of the pair relative to the detector, as
well as slight off-center placement of the primary. 

\begin{figure}
\plotone{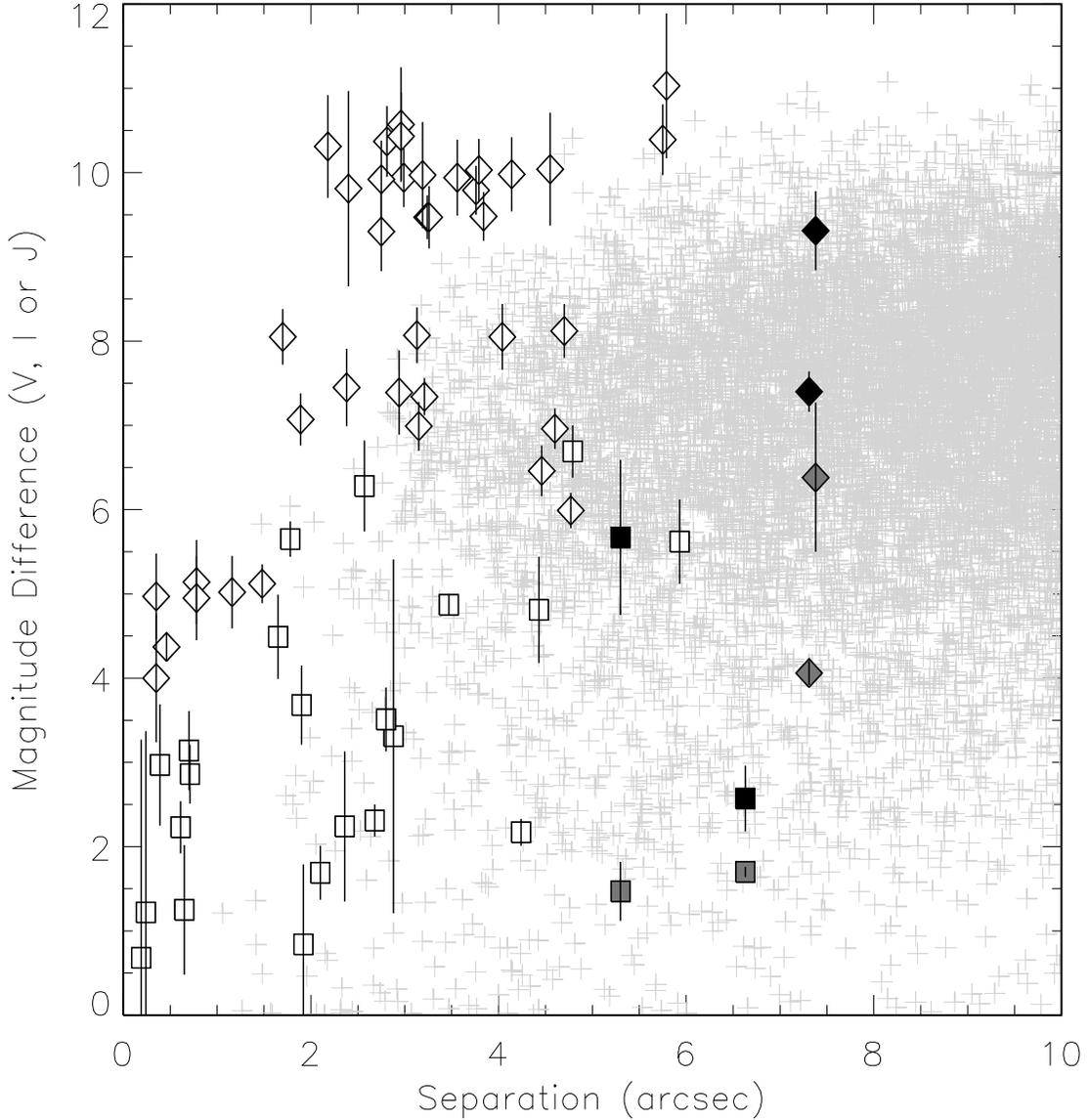}
\caption[2MASS and AO]{Magnitude difference versus separation from 2MASS and
AO.  AO observations are shown as squares (for previously-known pairs) or
diamonds (for new pairs), with error bars plotted for the photometry only.
The faint cloud of plus signs indicate all 2MASS pairs under 10$''$ in
separation, as described in the text.  Two known and two new AO pairs were
also found in the 2MASS data; these are shown as filled symbols, with 2MASS
$J$-band magnitude differences in grey and AO magnitude differences in black.
$V$-band magnitude differences for the two known pairs with AO and 2MASS data
were very similar to the AO $\triangle$m values, so are not plotted.}
\label{fig:jplot2}
\end{figure}

In order to get an idea of the separation/$\triangle$m limits of 2MASS, the
point-source catalog was searched for sources in the magnitude range 5.5 
$\le J \le$ 8.0, corresponding to the approximate 2MASS $J$-magnitude range
for the AO targets in this project.
This yielded 99,656 sources.
All sources within 10$''$ of these ``primaries'' were then extracted from the
catalog, yielding 9,657 companions.
A plot of separation versus $\triangle J$ is shown in Figure~\ref{fig:jplot2}.
As seen in this figure, very few close pairs are detected (only 17 pairs with
separations in the range $1'' < \rho < 2''$ and an additional 70 in
the range $2'' < \rho < 3''$).
For the main body of companions, maximum $\triangle J$ increases to perhaps
6-6.5 mag at 3$''$, 8-8.8 mag at 4$''$, and 9.5-10 mag at 7\farcs5.

\noindent {\it AO Single Stars}:
A Vizier search of the regions around each of the AO single stars found only
six 2MASS companions within 7\farcs5 of their respective primaries: HD~25638
(6\farcs2), HD~52266 (7\farcs1), HD~151515 (6\farcs2), HD~152219 (7\farcs2),
HD~163892 (6\farcs4), and HD~210809 (5\farcs6).
A further check of their orientations showed that all of these companions fell
either outside the observing window or so close to the edge of the field as to
not be measurable.
The companion to HD~25638 is known; see \S~\ref{unresolved_notes}.

\noindent {\it Known Pairs}:
A search of the regions around these primaries found six 2MASS companions
within 7\farcs5.
Two of these (HD~48279 and 152408) correspond to known companions also measured
by AO and are listed in Table~\ref{tbl:2mass}, which lists relative astrometry
and differential photometry in the 2MASS $J$, $H$, and $K_\text{S}$ bands.
The remaining four, including one known companion and three objects much wider
than the known secondaries, all fall outside the AO observing window.
Note that the companion to HD~152408 is very red
($\triangle m_{I} > \triangle m_{K_\text{S}}$), perhaps indicating that the
companion still possesses a luminous disk (like the case of the IR-bright
companion of $\sigma$~Ori~AB).

\noindent {\it New Pairs}:
The search around these objects found five 2MASS companions within
7$''$\llap.5.
Two of these (HD~156212 and 201345) correspond to the newly discovered AO
companions and are listed in Table~\ref{tbl:2mass}.
The other three --- two of which are known companions --- are all wide.

\section{DISCUSSION}
\label{sec:discussion}

\begin{figure}
\plotone{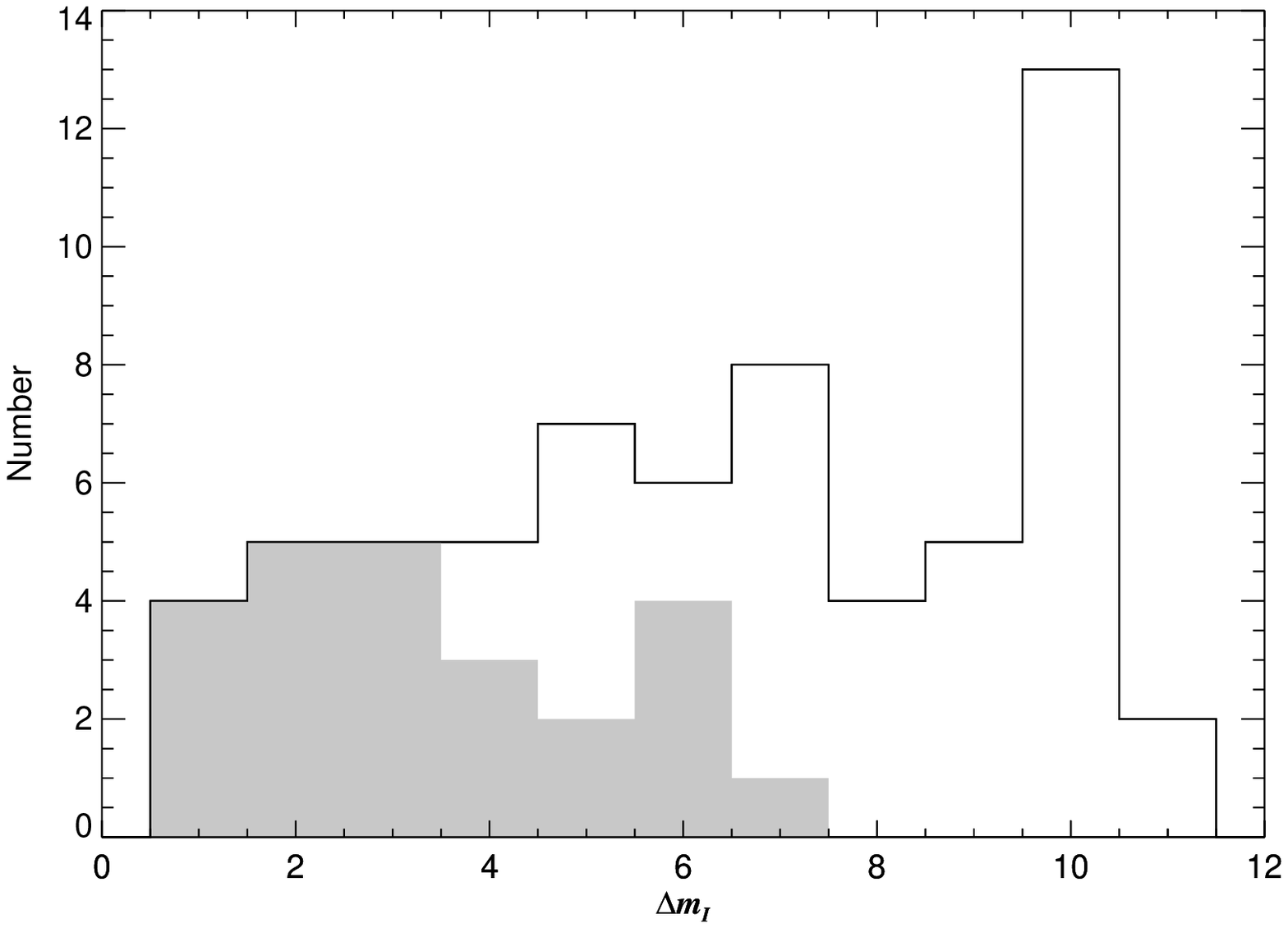}
\caption[Number vs. dI]{Number of companions vs. differential magnitude at
$I$-band.  Each bin represents a magnitude difference of 1 mag.  The shaded
region represents previously known companions while the solid line histogram
shows the total number of companions (known plus new).
}
\label{fig:hdi}
\end{figure}

Our AO survey of the O-stars has revealed a large number of new and generally
faint companions.
We show in Figure~\ref{fig:hdi} the distribution of numbers of companions
as a function of magnitude difference $\triangle I$ (in 1 mag bins), and the
distribution peaks near $\triangle I = 10$~mag.
It is certainly possible that some of these faint companions are low mass,
gravitationally bound stars.
For example, \citet{1997AJ....113.1733H} made a deep $I$-band survey of the
Orion cluster centered on the Orion Trapezium, and she found examples of still
embedded pre-main sequence stars with magnitude differences as large as
$\triangle I = 15$~mag compared to the bright O-stars in the Trapezium.
On the other hand, her survey also demonstrates that there will be many cluster
stars that appear projected on the sky at positions near O-stars but are not
orbitally bound to them.

We used the $I$-band star counts from \citet{1997AJ....113.1733H} to model the
possible confusion limit caused by nearby cluster stars that we might expect to
find in our observations.
The cumulative number distribution of stars brighter than $I$ in her sample
(1551 stars with measured $I$ magnitudes) is well matched by a power law,
\begin{equation}
N(<I) = 0.063~ 10^{\gamma I},
\end{equation}
where the exponent is $\gamma = 0.27$ (valid down to $I=16$ mag).
The total surface density of stars in the central part of cluster where the
O-stars reside is $10^{5.4}$ stars per square degree (see Figure~4a in
\citealt{1997AJ....113.1733H}) or 1.9 stars per $10\arcsec \times 10\arcsec$
field of view (= AEOS FOV).
Thus, the cumulative distribution of cluster background stars per AEOS FOV is
approximately given by
\begin{equation}
N(<I) = 7.8\times 10^{-5} ~ 10^{\gamma I}.
\end{equation}

If we suppose that all O-stars have a color index of $V-I=-0.39$
\citep{1994MNRAS.270..229W}, the extinction to Orion is approximately
$A_I=0.25$ \citep{1997AJ....113.1733H,1999PASP..111...63F}, and the distance
to Orion is 470~pc  \citep{1997AJ....113.1733H}, then the numbers per AEOS FOV
can be recast in terms of an O-star absolute magnitude $M_{V}$ and
magnitude difference $\triangle I$ as
\begin{equation}
N(<I) = 0.023~ 10^{\gamma (\triangle I + M_{V})}.
\end{equation}
If we further assume that the same relationship holds for other O-stars in
more distant clusters, then the number of cluster stars appearing in the FOV
will scale as distance $d$ squared.
Let $\eta$ represent the predicted number of cluster background stars per AEOS
FOV.
The relation between the limiting magnitude difference $\triangle I$
associated with $\eta$ is then
\begin{equation}
\triangle I = 3.6 -{2\over\gamma}\log d - M_{V}+{1\over\gamma}\log \eta.
\label{eq:eta}
\end{equation}
Thus, we expect that background confusion with cluster stars will increase
(smaller $\triangle I$) with increasing distance and with lower luminosity
O-stars.

We applied this relation to estimate a limiting $\triangle I$ for each cluster
and association star in our sample in order to separate the probable physical
companions from the faint optical companions caused by background confusion.
We used distances from \citet{MasonEtAl} and the spectral type -- absolute
magnitude calibration from \citet{2005AaA...436.1049M}.
The Orion cluster richness, extinction, and age may not be representative of
the full sample of cluster O-stars in our survey, and so the normalization of
the $\triangle I$ relation is somewhat uncertain.
We note that large numbers of companions begin to appear near
$\triangle I = 8$ mag (see Figure~\ref{fig:hdi}), and for typical O-star
distances ($d=1$ to 2 kpc) and magnitudes ($M_{V}= -4$ to $-6$) this
corresponds to a background density in the Orion model of $\eta = 1$ star per
AEOS FOV.
Thus, we adopted this value of $\eta$ in equation~\ref{eq:eta} to identify the
probable physical companions among the cluster and association stars.

The background limits for the field and runaway stars depend instead upon on
the stellar number density of Galactic background stars as a function of
magnitude and the direction of a given target.
Here we adopted the cumulative star counts as a function of apparent $I$
magnitude and Galactic coordinates from the Besan\c{c}on model of the Galaxy
\citep{2003AaA...409..523R}.
We created a cumulative star counts versus $\triangle I$ relation for each of
these stars and found a limiting magnitude for a specified value of 0.01 stars
per AEOS FOV.
We set this conservative limit based on numerical experiments for the set of
runaway stars with observed faint companions.
None of these companions can be physical companions since any such wide binary
would be disrupted at the time the runaway star was ejected (by close
gravitational encounters or a supernova in a binary).

The results of this exercise to discern the probable physical companions are
summarized in Table \ref{tbl:sum}.
The targets were placed in categories of cluster and association stars, field
stars, and runaway stars according to the criteria adopted by \citet{MasonEtAl}
(with some revisions from the subsequent work on field O-stars by
\citealt{2004AaA...425..937D, 2005AaA...437..247D}).
We see that after rejection of the possible background stars there is still a
high frequency ($37\%$) of detected companions among the cluster and
association stars in our survey.
In sharp contrast, there are very few such companions among the field and
runaway stars.
The only field star with a probable companion is HD~48279, and
\citet{2004AaA...425..937D} show that this star is $28\arcsec$ from a cluster
identified in 2MASS images, which suggests that this O-star may be a member of
a young, emerging cluster.
The only runaway star with a candidate companion is HD~34078 = AE~Aur, a star
ejected from the Orion association \citep{2004MNRAS.350..615G}. 
There is another optical companion, SEI~136, at a separation of $8\farcs4$
\citep{MasonEtAl}, and we suspect that both the companion we found and SEI~136
form a visual system that appears by a chance alignment with AE~Aur at this
time in its trajectory across the sky.

Our results confirm the trends found by \citet{MasonEtAl} that binary O-stars
are common among O-stars found near their birthplaces and are rare among
O-stars ejected from clusters.
These trends are consistent with current models for the ejection of runaways
and the formation of massive stars \citep{2007ARAaA..45..481Z}.
High velocity O-stars were probably ejected by close encounters between
binaries in dense clusters and by supernova explosions in close binaries
\citep{2001AaA...365...49H}, and their ejection velocities generally exceed the
escape velocity binding any wide, orbiting companion.
Many massive stars are probably born in high density stellar environments where
binaries may be formed through three-body interactions
\citep{2002MNRAS.336..705B} and through the interactions with large
proto-stellar accretion disks \citep{2007ApJ...661L.183M}.
Even in lower density environments, binaries may form through disk
fragmentation and subsequent gas accretion \citep{2007arXiv0709.4252K}.
Our results add to growing body of evidence that the formation of binaries is
closely linked to the formation of massive stars and the deposition of the
angular momentum reservoir of the natal cloud.

\acknowledgements

The United States Air Force provided the telescope time, on-site support, 
and 80\% of the research funds for this Air Force Office of Scientific
Research and National Science Foundation (NSF) jointly sponsored research 
under NSF grant number AST-0088498.
Additional support was provided by NSF grants AST-0506573 and AST-0606861.
LCR was funded by AFRL/DE (Contract Number F29601-00-D-0204).
We thank the numerous staff members of the Maui Space Surveillance System who
helped make this data possible.
Thanks as well to Gary Wycoff of USNO, who `mined' the 2MASS Catalog for
additional photometry and astrometry of these systems.
This research has made use of the {\it Washington Double Star Catalog},
maintained at the U.S. Naval Observatory, and of Aladin, Simbad, and Vizier,
operated at CDS, Strasbourg, France.
This publication also makes use of data products from the Two Micron All Sky
Survey, which is a joint project of the University of Massachusetts and 
the Infrared Processing and Analysis Center/California Institute of Technology,
funded by the National Aeronautics and Space Administration and the National 
Science Foundation.

% Revised by Gies 12-27-2007
\begin{deluxetable}{crrl@{~}r@{~}lccccc}
\tabletypesize{\small}
\tablecolumns{11}
\tablewidth{0pc}
\tablecaption{Survey Measurements\label{tbl:merge}}
\tablehead{
\multicolumn{1}{c}{} &
\multicolumn{1}{c}{} &
\multicolumn{1}{c}{} &
\multicolumn{3}{c}{Discoverer} &
\multicolumn{1}{c}{Epoch} &
\multicolumn{1}{c}{P.A.} &
\multicolumn{1}{c}{Sep.} &
\multicolumn{1}{c}{$\Delta m_I$} &
\multicolumn{1}{c}{Spec.} \\
\multicolumn{1}{c}{WDS} &
\multicolumn{1}{c}{HD} &
\multicolumn{1}{c}{HIP} &
\multicolumn{3}{c}{Designation} &
\multicolumn{1}{c}{(BY)} &
\multicolumn{1}{c}{($^{\circ}$)} &
\multicolumn{1}{c}{($''$)} &
\multicolumn{1}{c}{(mag)} &  
\multicolumn{1}{c}{Status\tablenotemark{c}} \\
\multicolumn{1}{c}{(1)} &
\multicolumn{1}{c}{(2)} &
\multicolumn{1}{c}{(3)} &
\multicolumn{3}{c}{(4)} &
\multicolumn{1}{c}{(5)} &
\multicolumn{1}{c}{(6)} &
\multicolumn{1}{c}{(7)} &
\multicolumn{1}{c}{(8)} &
\multicolumn{1}{c}{(9)} \\
}
\startdata
00061$+$6341 &    108 &    505 & TRN &    7 &      & 2001.7450 & \phn\phn5Z/176N\tablenotemark{a} & 3.24 & \phm{1}9.47$\pm$0.26 & C \\
00177$+$5126 &   1337 &   1415 & TRN &    8 &      & 2001.7341 &    \phn22Z/186N\tablenotemark{a} & 2.75 & \phm{1}9.30$\pm$0.47 & V \\
02158$+$5600 &  13745 &  10541 & TRN &    9 &      & 2004.7924 &                           \phn29 & 4.55 &       10.04$\pm$0.67 & V \\
02229$+$4129 &  14633 &  11099 & TRN &   10 &      & 2001.7451 &    \phn97Z/214N\tablenotemark{a} & 2.18 &       10.31$\pm$0.61 & V \\
02327$+$6127 &  15558 &  11832 & TRN &   11 & AF   & 2001.7450 &                              261 & 4.45 & \phm{1}6.59$\pm$0.19 & V \\
             &        &        &     &      &      & 2002.6845 &                              261 & 4.48 & \phm{1}6.38$\pm$0.40 & \\
             &        &        &     &      &      & 2005.6962 &                              262 & 4.44 & \phm{1}6.42$\pm$0.30 & \\
             &        &        & TRN &   11 & AG   & 2001.7450 &                              221 & 2.99 & \phm{1}9.94$\pm$0.35 & \\
             &        &        & TRN &   11 & AH   & 2001.7450 &                              211 & 5.79 &       11.03$\pm$0.86 & \\
02407$+$6117 &  16429 &  12495 & TRN &   12 & AD   & 2005.6963 &                              112 & 2.94 & \phm{1}7.39$\pm$0.22 & V \\
02511$+$6025 &  17505 &  13296 & STF &  306 & AB   & 2001.7450 &                           \phn91 & 2.09 & \phm{1}1.71$\pm$0.26 & V \\
             &        &        &     &      &      & 2002.6845 &                           \phn89 & 2.11 & \phm{1}1.71$\pm$0.37 & \\
             &        &        &     &      &      & 2004.9648 &                           \phn92 & 2.10 & \phm{1}1.64$\pm$0.37 & \\
             &        &        & TRN &   13 & AH   & 2001.7450 &                              142 & 4.59 & \phm{1}6.72$\pm$0.21 & \\
             &        &        &     &      &      & 2002.6845 &                              140 & 4.62 & \phm{1}7.01$\pm$0.27 & \\
             &        &        &     &      &      & 2004.9648 &                              143 & 4.59 & \phm{1}7.15$\pm$0.22 & \\
02594$+$6034 &  18326 &  13924 & TRN &   14 &      & 2001.7451 &       223Z/353N\tablenotemark{a} & 2.38 & \phm{1}7.45$\pm$0.46 & V \\
03141$+$5934 &  19820 &  15063 & TRN &   15 &      & 2001.7452 &    \phn81Z/209N\tablenotemark{a} & 2.81 &       10.37$\pm$0.42 & V \\
03556$+$5238 &  24431 &  18370 & HDS &  494 &      & 2001.7455 &                              176 & 0.71 & \phm{1}2.86$\pm$0.35 & C \\
03590$+$3548 &  24912 &  18614 & TRN &   16 &      & 2001.7342 &       141Z/236N\tablenotemark{a} & 2.40 & \phm{1}9.81$\pm$1.16 & C \\
05163$+$3419 &  34078 &  24575 & TRN &   17 & Aa   & 2004.7873 &                              171 & 0.35 & \phm{1}4.97$\pm$0.51 & C \\
05207$+$3726 &  34656 &  24957 & TRN &   18 & Aa   & 2002.7257 &                              279 & 0.35 & \phm{1}3.77$\pm$0.33 & C \\
             &        &        &     &      &      & 2003.7360 &                              280 & 0.35 & \phm{1}4.09$\pm$1.10 & \\
             &        &        &     &      &      & 2003.8917 &                              280 & 0.34 & \phm{1}4.13$\pm$0.84 & \\
             &        &        & TRN &   18 & AB   & 2002.7257 &                           \phn48 & 1.88 & \phm{1}6.98$\pm$0.24 & \\
             &        &        &     &      &      & 2003.7360 &                           \phn47 & 1.90 & \phm{1}6.94$\pm$0.25 & \\
             &        &        &     &      &      & 2003.8917 &                           \phn47 & 1.89 & \phm{1}7.28$\pm$0.43 & \\
             &        &        & TRN &   18 & AC   & 2003.7360 &                              126 & 2.96 &       10.57$\pm$0.68 & \\
             &        &        & TRN &   18 & AD   & 2003.7360 &                              247 & 5.75 &       10.39$\pm$0.42 & \\
05297$+$3523 &  35921 &  25733 & HU  &  217 &      & 2004.7873 &                              253 & 0.61 & \phm{1}2.23$\pm$0.31 & V \\
05351$+$0956 &  36861 &        & STF &  738 & AB   & 2001.7428 &     297Z/356N\tablenotemark{a,b} & 4.25 & \phm{1}2.29$\pm$0.16 & C \\
             &        &        &     &      &      & 2001.7456 &                           \phn41 & 4.23 & \phm{1}2.04$\pm$0.15 & \\
05353$-$0523 &  37022 &  26221 & STF &  748 & Ca,F & 2001.7456 &                              118 & 4.39 & \phm{1}4.89$\pm$1.06 & V \\
             &        &        &     &      &      & 2003.0045 &                              120 & 4.46 & \phm{1}4.73$\pm$0.19 & \\
05354$-$0525 &  37041 &  26235 & CHR &  249 & Aa   & 2004.8666 &                              292 & 0.39 & \phm{1}2.97$\pm$0.72 & V \\
05387$-$0236 &  37468 &  26549 & BU  & 1032 & AB   & 2001.7456 &                              106 & 0.24 & \phm{1}1.22$\pm$2.15 & V \\
             &        &        & TRN &   19 & AF   & 2001.7456 &                           \phn18 & 3.13 & \phm{1}8.07$\pm$0.33 & \\
05407$-$0157 &  37742 &        & STF &  774 & Aa-B & 2001.7455 &                              160 & 2.36 & \phm{1}2.24$\pm$0.89 & C \\
06319$+$0457 &  46150 &  31130 & GAN &    3 & AB   & 2001.0989 &                              284 & 3.47 & \phm{1}4.87$\pm$0.13 & C \\
06322$+$0450 &  46223 &  31149 & TRN &   20 &      & 2001.8661 &       314Z/264N\tablenotemark{a} & 0.46 & \phm{1}4.37$\pm$0.17 & C \\
06364$+$0605 &  46966 &  31567 & TRN &   21 &      & 2002.8023 &                              210 & 3.19 & \phm{1}9.97$\pm$0.63 & C \\
06374$+$0608 &  47129 &  31646 & TRN &   22 & AB   & 2002.2404 &                              240 & 1.15 & \phm{1}5.10$\pm$0.43 & V \\
             &        &        &     &      &      & 2005.0144 &                              251 & 1.16 & \phm{1}4.94$\pm$0.22 & \\
             &        &        & TRN &   22 & AC   & 2005.0144 &                              203 & 0.78 & \phm{1}5.14$\pm$0.22 & \\
06386$+$0137 &  47432 &  31766 & TRN &   23 &      & 2005.0134 &                              206 & 0.78 & \phm{1}4.95$\pm$1.31 & C \\
06410$+$0954 &  47839 &  31978 & STF &  950 & Aa-B & 2001.7457 &                              202 & 2.88 & \phm{1}3.31$\pm$2.11 & V \\
06427$+$0143 &  48279 &  32137 & STF &  956 & AB   & 2004.0478 &                              199 & 6.63 & \phm{1}2.57$\pm$0.39 & C \\
08392$-$4025 &  73882 &  42433 & B   & 1623 &      & 2004.1190 &                              256 & 0.65 & \phm{1}1.25$\pm$0.77 & V \\
16550$-$4109 & 152408 &  82775 & I   &  576 &      & 2001.4030 &       260Z/261N\tablenotemark{a} & 5.30 & \phm{1}5.67$\pm$0.92 & C \\
17065$-$3527 & 154368 &  83706 & B   &  894 &      & 2002.6730 &                              357 & 2.57 & \phm{1}6.28$\pm$0.54 & V \\
17158$-$3344 & 155889 &  84444 & SEE &  322 &      & 2003.6041 &                              285 & 0.19 & \phm{1}0.68$\pm$2.59 & C \\
17175$-$2746 & 156212 &  84588 & TRN &   24 & AB   & 2001.4959 &       206Z/204N\tablenotemark{a} & 4.04 & \phm{1}8.05$\pm$0.39 & C \\
             &        &        & TRN &   24 & AC   & 2001.4959 &       145Z/144N\tablenotemark{a} & 7.31 & \phm{1}7.40$\pm$0.24 & \\
17347$-$3235 & 159176 &  86011 & HDS & 2480 & Ab   & 2003.3280 &                           \phn59 & 0.70 & \phm{1}3.14$\pm$0.47 & V \\
17595$-$3601 & 163758 &        & TRN &   25 &      & 2001.4960 &    107Z/\phn93N\tablenotemark{a} & 1.70 & \phm{1}8.05$\pm$0.33 & C \\
18024$-$2302 & 164492 &  88333 & TRN &   26 & AH   & 2002.4516 &                              342 & 1.48 & \phm{1}5.12$\pm$0.23 & V \\
18026$-$2415 & 164536 &        & RST & 3149 & AB   & 2002.4627 &                           \phn62 & 1.65 & \phm{1}4.49$\pm$12\tablenotemark{d} & C \\
18061$-$2412 & 165246 &        & B   &  376 &      & 2002.4819 &                           \phn98 & 1.90 & \phm{1}3.68$\pm$0.47 & V \\
18152$-$2023 & 167263 &        & BU  &  286 & AB   & 2005.6545 &                              216 & 5.93 & \phm{1}5.62$\pm$0.98 & V \\
18181$-$1215 & 167971 &  89681 & TRN &   27 &      & 2001.7447 & \phn80Z/\phn39N\tablenotemark{a} & 4.70 & \phm{1}8.12$\pm$0.32 & V \\
20035$+$3601 & 190429 &  98753 & STF & 2624 & Aa-B & 2001.7337 &                              172 & 1.92 & \phm{1}0.84$\pm$0.95 & C \\
20181$+$4044 & 193322 & 100069 & STF & 2666 & Aa-B & 2001.6683 &                              244 & 2.69 & \phm{1}2.51$\pm$0.12 & V \\
             &        &        &     &      &      & 2001.7364 &                              243 & 2.67 & \phm{1}2.11$\pm$0.26 & \\
20191$+$3916 & 193514 & 100173 & TRN &   28 & AB   & 2001.7367 &                           \phn78 & 4.75 & \phm{1}5.80$\pm$0.17 & C \\
             &        &        &     &      &      & 2002.6730 &                           \phn79 & 4.79 & \phm{1}6.18$\pm$0.25 & \\
             &        &        & TRN &   28 & AC   & 2001.7367 &                              157 & 3.20 & \phm{1}7.23$\pm$0.28 & \\
             &        &        &     &      &      & 2002.6730 &                              158 & 3.22 & \phm{1}7.44$\pm$0.16 & \\
20205$+$4351 & 193793 & 100287 & BU  & 1207 & AB   & 2001.3400 &                              209 & 4.77 & \phm{1}8.44$\pm$0.31 & V \\
             &        &        &     &      &      & 2001.7337 &                              209 & 4.77 & \phm{1}6.75$\pm$0.28 & \\
             &        &        &     &      &      & 2001.7363 &                              209 & 4.77 & \phm{1}6.75$\pm$12\tablenotemark{d} & \\
             &        &        &     &      &      & 2001.7447 &         \nodata\tablenotemark{b} & 4.79 & \phm{1}6.43$\pm$0.53 & \\
             &        &        &     &      &      & 2002.6731 &                              210 & 4.81 & \phm{1}6.58$\pm$0.13 & \\
             &        &        &     &      &      & 2002.6731 &                              210 & 4.81 & \phm{1}6.93$\pm$0.12 & \\
             &        &        & TRN &   29 & AC   & 2001.7363 &                              202 & 2.29 & \phm{1}9.62$\pm$0.35 & \\
             &        &        &     &      &      & 2001.7447 &         \nodata\tablenotemark{b} & 3.23 & \phm{1}9.32$\pm$0.38 & \\
20566$+$4455 & 199579 & 103371 & TRN &   30 &      & 2001.7338 &    \phn30Z/205N\tablenotemark{a} & 3.76 & \phm{1}9.79$\pm$0.29 & V \\
21079$+$3324 & 201345 & 104316 & TRN &   31 &      & 2001.7367 &       342Z/218N\tablenotemark{a} & 7.38 & \phm{1}9.31$\pm$0.47 & C \\
21185$+$4357 & 203064 & 105186 & TRN &   32 &      & 2001.7337 &    \phn81Z/219N\tablenotemark{a} & 3.84 & \phm{1}9.48$\pm$0.29 & C \\
21390$+$5729 & 206267 & 106886 & BU  & 1143 & AB   & 2001.7338 &                              317 & 1.78 & \phm{1}5.65$\pm$0.21 & V \\
21449$+$6228 & 207198 & 107374 & TRN &   33 & AC   & 2001.7364 &    \phn75Z/207N\tablenotemark{a} & 2.96 &       10.43$\pm$0.52 & C \\
22021$+$5800 & 209481 & 108772 & TRN &   34 &      & 2001.7341 &       358Z/208N\tablenotemark{a} & 2.75 & \phm{1}9.92$\pm$0.46 & V \\
22051$+$6217 & 209975 & 109017 & TRN &   35 & AD   & 2001.7367 &       341Z/171N\tablenotemark{a} & 4.14 & \phm{1}9.98$\pm$0.44 & C \\
             &        &        & TRN &   35 & AE   & 2001.7367 &       344Z/175N\tablenotemark{a} & 3.79 &       10.03$\pm$0.37 & \\
22393$+$3903 & 214680 & 111841 & TRN &   36 & AC   & 2001.7339 &    \phn85Z/222N\tablenotemark{a} & 3.56 & \phm{1}9.94$\pm$0.45 & C \\
22568$+$6244 & 217086 & 113306 & MLR &  266 & AB   & 2001.7369 &                              353 & 2.79 & \phm{1}3.41$\pm$0.27 & C \\
             &        &        &     &      &      & 2002.6843 &                              354 & 2.83 & \phm{1}3.34$\pm$0.48 & \\
             &        &        &     &      &      & 2004.7598 &                              354 & 2.78 & \phm{1}3.79$\pm$0.58 & \\
             &        &        & TRN &   37 & AC   & 2001.7369 &                              164 & 3.15 & \phm{1}7.14$\pm$0.29 & \\
             &        &        &     &      &      & 2002.6843 &                              165 & 3.17 & \phm{1}6.79$\pm$12\tablenotemark{d} & \\
             &        &        &     &      &      & 2004.7598 &                              165 & 3.12 & \phm{1}7.03$\pm$0.34 & \\
\enddata
\tablenotetext{a}{Image derotator in an unknown state: Z = zenith up and N = north up. Confirmation of the pair will 
clear up this residual ambiguity.}
\tablenotetext{b}{Image derotator in an unknown state. Neither Zenith up or North up match coordinates of known object in field.}
\tablenotetext{c}{Codes are: {\bf V} - one or more additional spectroscopic companions indicated, or {\bf C} - radial velocity constant.}
\tablenotetext{d}{This is a lower limit for the errors, the actual errors are typically much larger.}
\end{deluxetable}

\begin{deluxetable}{ll@{~}r@{~}lrrrrrrr}
\tabletypesize{\small}
%\tablenum{3}
\rotate
\tablewidth{0pt}
\tablecaption{Orbital Elements\label{tbl:orb_el}}
\tablehead{
\colhead{WDS Desig.} & 
\multicolumn{3}{c}{Discoverer} &
\colhead{Period} &
\colhead{Semimajor} & 
\colhead{Inclination} &
\colhead{Longitude} &
\colhead{Epoch of} &
\colhead{Eccentricity} &
\colhead{Longitude of} \\
\colhead{~} & 
\multicolumn{3}{c}{Designation} &
\colhead{~} &
\colhead{Axis} & 
\colhead{~} &
\colhead{of Node} &
\colhead{Periastron} &
\colhead{~} &
\colhead{Periastron} \\
\colhead{$\alpha \delta$ (2000)} &
\multicolumn{3}{c}{~} & 
\colhead{$P$ (yr)} & 
\colhead{$a$ ($''$)} & 
\colhead{$i$ ($\circ$)} &
\colhead{$\Omega$ ($\circ$)} &
\colhead{$T_{o}$ (yr)} &
\colhead{$e$} & 
\colhead{$\omega$ ($\circ$)} \\
}
\startdata
05387$-$0236 & BU  & 1032 & AB &    156.7~~     &      0.2662~~~       &    159.7~~~~~    &     121.7~~~~    &    1999.5~~~~    &      0.0515~~~~~       &         8.7~~~~~~~    \\
             &     &      &    & $\pm$3.0~~     & $\pm$0.0021~~~       & $\pm$3.7~~~~~    &  $\pm$9.6~~~~    & $\pm$10.2~~~~    & $\pm$0.0080~~~~~       &   $\pm$16.9~~~~~~~    \\
             &     &      &    &          ~     &              ~       &          ~~      &       ~~~~       &    ~~~~          &            ~~~~~       &            ~~~~~~~    \\
05387$-$0236 & BU  & 1032 & AB\tablenotemark{a} &    155.3~~     &      0.2642~~~       &    160.4~~~~~    &     136.\phn~~~~ &    1997.\phn~~~~ &      0.051\phn~~~~~    &       18.\phn~~~~~~~  \\
             &     &      &    & $\pm$7.5~~     & $\pm$0.0052~~~       & $\pm$7.2~~~~~    & $\pm$25.\phn~~~~ & $\pm$24.\phn~~~~ & $\pm$0.015\phn~~~~~    &  $\pm$37.\phn~~~~~~~  \\
             &     &      &    &          ~     &              ~       &          ~~      &      ~~~~        &         ~~~~     &             ~~~~~      &              ~~~~~~~  \\
17158$-$3344 & SEE &  322 &    &    536.\phn~~  &      0.35\phm{88}~~~ &    111.\phn~~~~~ &     167.\phn~~~~ &    1870.\phn~~~~ &      0.64\phm{88}~~~~~ &      109.\phn~~~~~~~  \\
\enddata
\tablenotetext{a}{Elements from \cite{1996AJ....111..370H}.}
\end{deluxetable}

%\begin{deluxetable}{ll@{~}r@{~}l@{~~~~~}r@{~~~}l@{~~~~~}r@{~~~}l@{~~~~~}r@{~~~}l@{~~~~~}r@{~~~}l@{~~~~~}r@{~~~}l}
\begin{deluxetable}{cl@{~}r@{~}lc@{~~}cc@{~~}cc@{~~}cc@{~~}cc@{~~}c}
\tabletypesize{\small}
%\rotate
%\tablenum{3}
\tablewidth{0pt}
\tablecaption{Future Ephemerides (BY) for BU~1032 and SEE~322\label{tbl:ephem}}
\tablehead{
\colhead{WDS Desig.} & 
\multicolumn{3}{c}{Discoverer} &
\multicolumn{2}{c}{2008} & 
\multicolumn{2}{c}{2010} &
\multicolumn{2}{c}{2012} &
\multicolumn{2}{c}{2014} &
\multicolumn{2}{c}{2016} \\
\colhead{$\alpha,\delta$ (2000)} &
\multicolumn{3}{c}{Designation} & 
\multicolumn{2}{c}{$\theta$ ($^\circ$)~~$\rho$ ($''$)} &
\multicolumn{2}{c}{$\theta$ ($^\circ$)~~$\rho$ ($''$)} &
\multicolumn{2}{c}{$\theta$ ($^\circ$)~~$\rho$ ($''$)} &
\multicolumn{2}{c}{$\theta$ ($^\circ$)~~$\rho$ ($''$)} &
\multicolumn{2}{c}{$\theta$ ($^\circ$)~~$\rho$ ($''$)} \\
}
\startdata
05387$-$0236 & BU  & 1032 & AB   & \phn93.0 & 0.249 & \phn88.0 & 0.249 & \phn83.1 & 0.248 & \phn78.1 & 0.247 &    \phn73.1 & 0.246 \\
17158$-$3344 & SEE &  322 &      &    281.7 & 0.184 &    280.4 & 0.184 &    279.0 & 0.183 &    277.7 & 0.183 &       276.3 & 0.183 \\
\enddata
\end{deluxetable}

% Revised by Gies 12-27-2007
\begin{deluxetable}{rrccc}
\tablecolumns{5}
%\tablenum{5}
\tablewidth{0pc}
\tablecaption{Single Star FWHM measures\label{tbl:sing}}
\tablehead{
\multicolumn{1}{c}{} &
\multicolumn{1}{c}{} &
\multicolumn{1}{c}{Epoch} &
\multicolumn{1}{c}{FWHM} &
\multicolumn{1}{c}{Spec.} \\
\multicolumn{1}{c}{HD} &
\multicolumn{1}{c}{HIP} &
\multicolumn{1}{c}{(BY)} &
\multicolumn{1}{c}{($''$)} &
\multicolumn{1}{c}{Status\tablenotemark{a}} \\
}
\startdata
 14947 &  11394 & 2001.7455 & 0.12 & C  \\
 15137 &  11473 & 2004.7923 & 0.18 & V  \\
 25638 &  19272 & 2004.8665 & 0.12 & C  \\
 30614 &  22783 & 2004.8665 & 0.12 & C  \\
 36879 &  26272 & 2003.8944 & 0.11 & C  \\
 37042 &        & 2004.8667 & 0.12 & V  \\
 37043 &  26241 & 2001.7455 & 0.08 & V  \\
 37366 &  26611 & 2005.0200 & 0.34 & V  \\
 39680 &  27941 & 2002.2404 & 0.22 & C  \\
 41161 &  28881 & 2001.8659 & 0.16 & C  \\
 42088 &  29216 & 2001.8660 & 0.21 & C  \\
 45314 &  30722 & 2001.0989 & 0.11 & V  \\
 46149 &  31128 & 2001.8660 & 0.36 & V  \\
 46966 &  31567 & 2002.2377 & 0.25 & C  \\
 48099 &  32067 & 2002.2405 & 0.13 & V  \\
 52266 &  33723 & 2001.9754 & 0.12 & V  \\
 53975 &  34297 & 2005.0196 & 0.12 & V  \\
 54662 &  34536 & 2002.0218 & 0.16 & V  \\
       &        & 2002.2404 & 0.12 &    \\
 55879 &  34999 & 2005.0170 & 0.19 & C  \\
 57061 &  35415 & 2004.0807 & 0.21 & V  \\
 57682 &  35707 & 2001.9290 & 0.45 & C  \\
 60848 &  37074 & 2001.0990 & 0.16 & C  \\
 66811 &  39429 & 2004.0861 & 0.48 & C  \\
 69648 &        & 2004.0451 & 0.26 & \nodata  \\
 75211 &  43125 & 2004.1432 & 0.23 & \nodata  \\
 75222 &  43158 & 2005.0197 & 0.49 & C  \\
 93521 &  52849 & 2001.0992 & 0.19 & C  \\
148546 &  80829 & 2005.3482 & 0.28 & C  \\
149404 &  81305 & 2001.3893 & 0.12 & V  \\
149757 &  81377 & 2002.2243 & 0.22 & C  \\
151003 &  82121 & 2004.2590 & 0.23 & V  \\
151515 &  82366 & 2004.3384 & 1.93 & C  \\
152003 &        & 2001.3893 & 0.12 & C  \\
152219 &        & 2004.5265 & 0.22 & V  \\
152314 &        & 2001.4029 & 0.17 & C  \\
153919 &  83499 & 2003.5247 & 0.16 & V  \\
155806 &  84401 & 2002.6730 & 0.12 & C  \\
157857 &  85331 & 2002.2383 & 0.22 & C  \\
162978 &  87706 & 2001.7446 & 0.12 & C  \\
163758 &        & 2001.5179 & 0.17 & C  \\
163800 &  88040 & 2001.4960 & 0.13 & \nodata  \\
       &        & 2001.5176 & 0.16 &    \\
163892 &  88085 & 2005.3404 & 0.21 & V  \\
164438 &  88297 & 2002.3204 & 0.16 & \nodata  \\
164794 &  88469 & 2002.6729 & 0.10 & V  \\
164816 &        & 2001.6709 & 0.14 & V  \\
       &        & 2002.4817 & 0.11 &    \\
165052 &  88581 & 2001.6711 & 0.14 & V  \\
165319 &  88652 & 2001.6711 & 0.16 & C  \\
167771 &  89630 & 2002.4843 & 0.20 & V  \\
       &        & 2002.4950 & 0.19 &    \\
175876 &  93118 & 2001.7447 & 0.14 & C  \\
186980 &  97280 & 2001.7366 & 0.19 & C  \\
188001 &  97796 & 2001.7338 & 0.12 & V  \\
       &        & 2001.7447 & 0.17 &    \\
188209 &  97757 & 2004.7646 & 0.08 & C  \\
190864 &  98976 & 2001.7365 & 0.14 & V  \\
190918 &  99002 & 2001.7421 & 0.18 & V  \\
191612 &  99308 & 2001.7365 & 0.13 & C  \\
191978 &  99439 & 2002.6786 & 0.33 & C  \\
192281 &  99580 & 2001.7365 & 0.12 & C  \\
192639 &  99768 & 2001.7366 & 0.17 & C  \\
193443 & 100146 & 2001.7366 & 0.19 & V  \\
195592 & 101186 & 2005.6549 & 0.19 & V  \\
198846 & 102999 & 2005.3843 & 0.12 & V  \\
202124 & 104695 & 2005.6549 & 0.19 & C  \\
203064 & 105186 & 2001.7342 & 0.12 & C  \\
       &        & 2001.7448 & 0.12 &    \\
207538 & 107598 & 2005.7180 & 0.22 & C  \\
210809 & 109562 & 2001.7369 & 0.21 & C  \\
210839 & 109556 & 2001.7342 & 0.16 & C  \\
214680 & 111841 & 2001.7369 & 0.19 & C  \\
216898 & 113218 & 2001.7451 & 0.10 & \nodata  \\
218915 & 114482 & 2004.7597 & 0.18 & C  \\
\enddata
\tablenotetext{a}{Codes are: {\bf V} - one or more additional spectroscopic
companions indicated, or {\bf C} - radial velocity constant.}
\end{deluxetable}

\begin{deluxetable}{l@{~}r@{~}lrlrrccccccc}
%\tablenum{6}
\tablecolumns{12}
\tablewidth{0pc}
\tablecaption{Data Mining Results from 2MASS\label{tbl:2mass}}
\tablehead{
\multicolumn{3}{c}{Discoverer} &
\multicolumn{1}{c}{~HD} &
\multicolumn{1}{c}{~HIP} &
\multicolumn{1}{c}{WDS} &
\multicolumn{1}{c}{Epoch} &
\multicolumn{1}{c}{P.A.} &
\multicolumn{1}{c}{Sep.} &
\multicolumn{1}{c}{$\Delta\text{m}_{J}$} &
\multicolumn{1}{c}{$\Delta\text{m}_{H}$} &
\multicolumn{1}{c}{$\Delta\text{m}_{K_{s}}$} \\
\multicolumn{3}{c}{Designation} &  & & &
\multicolumn{1}{c}{(BY)} &
\multicolumn{1}{c}{($^{\circ}$)} &
\multicolumn{1}{c}{($''$)} & 
\multicolumn{1}{c}{(mag)} &
\multicolumn{1}{c}{(mag)} &
\multicolumn{1}{c}{(mag)} \\
}
\startdata
STF &  956 & AB &  48279 &  32137 & 06427$+$0143 & 1999.87 &  194.8  & 6.55 & {\phm{$\pm$}}1.70 & {\phm{$\pm$}}1.63 & {\phm{$\pm$}}1.56 \\
    &      &    &        &        &              &         &         &      &         $\pm$0.06 &         $\pm$0.03 &         $\pm$0.04 \\
I   &  576 &    & 152408 &  82775 & 16550$-$4109 & 1999.36 &  264.2  & 4.72 & {\phm{$\pm$}}1.47 & {\phm{$\pm$}}1.71 & {\phm{$\pm$}}1.25 \\
    &      &    &        &        &              &         &         &      &         $\pm$0.35 &         $\pm$0.58 &         $\pm$0.44 \\
TRN &   24 & AC & 156212 &  84588 & 17175$-$2746 & 1998.53 &  147.6  & 7.32 & {\phm{$\pm$}}4.06 & {\phm{$\pm$}}5.38 & {\phm{$\pm$}}4.24 \\
    &      &    &        &        &              &         &         &      &                   &         $\pm$0.14 &                   \\ 
TRN &   31 &    & 201345 & 104316 & 21079$+$3324 & 1999.75 &  220.8  & 7.44 & {\phm{$\pm$}}6.38 & {\phm{$\pm$}}5.88 & {\phm{$\pm$}}5.69 \\
    &      &    &        &        &              &         &         &      &         $\pm$0.89 &         $\pm$0.45 &         $\pm$0.24 \\          
\enddata
\end{deluxetable}

\begin{deluxetable}{lcccc}
\tablecolumns{5}
\tablewidth{0pc}
\tablecaption{Companion Frequency and Environment\label{tbl:sum}}
\tablehead{
\multicolumn{1}{c}{Category} &
\multicolumn{1}{c}{Number} &
\multicolumn{1}{c}{Number} &
\multicolumn{1}{c}{Number} &
\multicolumn{1}{c}{Percentage} \\
& \multicolumn{1}{c}{of Stars} &
\multicolumn{1}{c}{w/ Comp.} &
\multicolumn{1}{c}{w/ Phys. Comp.} &
\multicolumn{1}{c}{of Phys. Comp.} \\
}
\startdata
Cluster/Association \dotfill &     83 &     41 &     31 & $37\pm7$  \\
Field               \dotfill & \phn 9 & \phn 2 & \phn 1 & $11\pm11$ \\
Runaway             \dotfill &     24 & \phn 8 & \phn 1 & $4\pm4$   \\
\enddata
\end{deluxetable}

\newcommand{\SortNoop}[1]{}


\begin{thebibliography}{}

\bibitem[Abt et~al., 1972]{1972AJ.....77..138A}
Abt, H.~A., Levy, S.~G., \& Gandet, T.~L. 1972,
\newblock \aj, 77, 138

\bibitem[Africano et~al., 1975]{1975AJ.....80..689A}
Africano, J.~L., Cobb, C.~L., Dunham, D.~W., Evans, D.~S., Fekel, F.~C., \&
  Vogt, S.~S. 1975,
\newblock \aj, 80, 689

\bibitem[Bate et~al., 2002]{2002MNRAS.336..705B}
Bate, M.~R., Bonnell, I.~A., \& Bromm, V. 2002,
\newblock \mnras, 336, 705

\bibitem[Bonnell \& Bate, 2005]{2005MNRAS.362..915B}
Bonnell, I.~A. \& Bate, M.~R. 2005,
\newblock \mnras, 362, 915

\bibitem[{\SortNoop{Bos}}{van den Bos}, 1928]{vandenBos1928}
{\SortNoop{Bos}}{van den Bos}, W.~H. 1928,
\newblock Ann. Leiden Obs., 14,
\newblock pt. 4

\bibitem[Boyajian et~al., 2005]{2005ApJ...621..978B}
Boyajian, T.~S. et~al. 2005,
\newblock \apj, 621, 978

\bibitem[Boyajian et~al., 2007a]{2007PASP..119..742B}
Boyajian, T.~S. et~al. 2007a,
\newblock \pasp, 119, 742

\bibitem[Boyajian et~al., 2007b]{2007ApJ...664.1121B}
Boyajian, T.~S. et~al. 2007b,
\newblock \apj, 664, 1121

\bibitem[{\SortNoop{Brummelaar}}{ten Brummelaar} et~al., 1998]{spie3353-391}
{\SortNoop{Brummelaar}}{ten Brummelaar}, T.~A., Hartkopf, W.~I., McAlister,
  H.~A., Mason, B.~D., Roberts, Jr., L.~C., \& Turner, N.~H. 1998,
\newblock in \procspie, Adaptive Optical System Technologies, ed. D.~Bonaccini
  \& R.~K. Tyson (Bellingham, WA: SPIE), 3353, 391

\bibitem[{\SortNoop{Brummelaar}}{ten Brummelaar} et~al., 1996]{diffmag1}
{\SortNoop{Brummelaar}}{ten Brummelaar}, T.~A., Mason, B.~D., Bagnuolo, Jr.,
  W.~G., Hartkopf, W.~I., McAlister, H.~A., \& Turner, N.~H. 1996,
\newblock \aj, 112, 1180

\bibitem[{\SortNoop{Brummelaar}}{ten Brummelaar} et~al., 2000]{diffmag2}
{\SortNoop{Brummelaar}}{ten Brummelaar}, T.~A., Mason, B.~D., McAlister, H.~A.,
  Roberts, Jr., L.~C., Turner, N.~H., Hartkopf, W.~I., \& Bagnuolo, Jr., W.~G.
  2000,
\newblock \aj, 119, 2403

\bibitem[Caballero, 2007]{2007AaA...466..917C}
Caballero, J.~A. 2007,
\newblock \aap, 466, 917

\bibitem[Caballero, 2008]{2008MNRAS.383..750C}
Caballero, J.~A. 2008,
\newblock \mnras, 383, 750

\bibitem[Cox, 2001]{APQ2001}
Cox, A.~N. 2001,
\newblock Allen's {A}strophysical {Q}uantities (4th ed.; New York, NY:
  Springer)

\bibitem[Cutri et~al., 2003]{2MASS}
Cutri, R.~M. et~al. 2003,
\newblock 2{MASS} {A}ll {S}ky {C}atalog of {P}oint {S}ources (Pasadena, CA:
  NASA/IPAC)

\bibitem[Fabricius \& Makarov, 2000]{2000AaA...356..141F}
Fabricius, C. \& Makarov, V.~V. 2000,
\newblock \aap, 356, 141

\bibitem[Fitzpatrick, 1999]{1999PASP..111...63F}
Fitzpatrick, E.~L. 1999,
\newblock \pasp, 111, 63

\bibitem[Gies \& Bolton, 1986]{1986ApJS...61..419G}
Gies, D.~R. \& Bolton, C.~T. 1986,
\newblock \apjs, 61, 419

\bibitem[Gualandris et~al., 2004]{2004MNRAS.350..615G}
Gualandris, A., {Portegies Zwart}, S., \& Eggleton, P.~P. 2004,
\newblock \mnras, 350, 615

\bibitem[Hartkopf et~al., 1996]{1996AJ....111..370H}
Hartkopf, W.~I., Mason, B.~D., \& McAlister, H.~A. 1996,
\newblock \aj, 111, 370

\bibitem[Hartkopf et~al., 2001]{2001AJ....122.3472H}
Hartkopf, W.~I., Mason, B.~D., \& Worley, C.~E. 2001,
\newblock \aj, 122, 3472

\bibitem[Hillenbrand, 1997]{1997AJ....113.1733H}
Hillenbrand, L.~A. 1997,
\newblock \aj, 113, 1733

\bibitem[Hillwig et~al., 2006]{2006ApJ...639.1069H}
Hillwig, T.~C., Gies, D.~R., Bagnuolo, Jr., W.~G., Huang, W., McSwain, M.~V.,
  \& Wingert, D.~W. 2006,
\newblock \apj, 639, 1069

\bibitem[Hinkley et~al., 2007]{2007ApJ...654..633H}
Hinkley, S. et~al. 2007,
\newblock \apj, 654, 633

\bibitem[Hoogerwerf et~al., 2001]{2001AaA...365...49H}
Hoogerwerf, R., {\SortNoop{Bruijne}}{de Bruijne}, J. H.~J., \&
  {\SortNoop{Zeeuw}}{de Zeeuw}, P.~T. 2001,
\newblock \aap, 365, 49

\bibitem[Hopmann, 1967]{1967MiWie..13...49H}
Hopmann, J. 1967,
\newblock Mitteilungen der Universitaets-Sternwarte Wien, 13, 49

\bibitem[Hummel et~al., 2000]{2000ApJ...540L..91H}
Hummel, C.~A., White, N.~M., Elias, II, N.~M., Hajian, A.~R., \& Nordgren,
  T.~E. 2000,
\newblock \apjl, 540, L91

\bibitem[Kratter et~al., 2007]{2007arXiv0709.4252K}
Kratter, K.~M., Matzner, C.~D., \& Krumholz, M.~R. 2007,
\newblock ArXiv e-prints, 0709.4252

\bibitem[Kraus et~al., 2007]{2007AaA...466..649K}
Kraus, S. et~al. 2007,
\newblock \aap, 466, 649

\bibitem[{\SortNoop{Loon}}{van Loon} \& Oliveira, 2003]{2003AaA...405L..33V}
{\SortNoop{Loon}}{van Loon}, J.~T. \& Oliveira, J.~M. 2003,
\newblock \aap, 405, L33

\bibitem[Ma\'{\i}z-Apell\'{a}niz et~al., 2004]{GOS}
Ma\'{\i}z-Apell\'{a}niz, J., Walborn, N.~R., Galu\'{e}, H.~A., \& Wei, L. 2004,
\newblock \apjs, 151, 103

\bibitem[Makidon et~al., 2005]{2005PASP..117..831M}
Makidon, R.~B., Sivaramakrishnan, A., Perrin, M.~D., Roberts~Jr., L.~C.,
  Oppenheimer, B.~R., Soummer, R., \& Graham, J.~R. 2005,
\newblock \pasp, 117, 831

\bibitem[Martins et~al., 2005]{2005AaA...436.1049M}
Martins, F., Schaerer, D., \& Hillier, D.~J. 2005,
\newblock \aap, 436, 1049

\bibitem[Mason et~al., 1998]{MasonEtAl}
Mason, B.~D., Hartkopf, W.~I., Gies, D.~R., Bagnuolo, Jr., W.~G.,
  {\SortNoop{Brummelaar}}{ten Brummelaar}, T.~A., \& McAlister, H.~A. 1998,
\newblock \aj, 115, 821

\bibitem[Mason et~al., 2001a]{2001AJ....121.3224M}
Mason, B.~D., Hartkopf, W.~I., Holdenried, E.~R., \& Rafferty, T.~J. 2001a,
\newblock \aj, 121, 3224

\bibitem[Mason et~al., 2004]{2004AJ....127..539M}
Mason, B.~D. et~al. 2004,
\newblock \aj, 127, 539

\bibitem[Mason et~al., 2001b]{WDS2001}
Mason, B.~D., Wycoff, G.~L., Hartkopf, W.~I., Douglass, G.~G., \& Worley, C.~E.
  2001b,
\newblock \aj, 122, 3466

\bibitem[Massey et~al., 1995]{1995AJ....110.2715M}
Massey, P., Armandroff, T.~E., Pyke, R., Patel, K., \& Wilson, C.~D. 1995,
\newblock \aj, 110, 2715

\bibitem[McSwain, 2003]{2003ApJ...595.1124M}
McSwain, M.~V. 2003,
\newblock \apj, 595, 1124

\bibitem[McSwain et~al., 2007]{2007ApJ...655..473M}
McSwain, M.~V., Boyajian, T.~S., Grundstrom, E.~D., \& Gies, D.~R. 2007,
\newblock \apj, 655, 473

\bibitem[Mdzinarishvili \& Chargeishvili, 2005]{2005AaA...431L...1M}
Mdzinarishvili, T.~G. \& Chargeishvili, K.~B. 2005,
\newblock \aap, 431, L1

\bibitem[Moeckel \& Bally, 2007]{2007ApJ...661L.183M}
Moeckel, N. \& Bally, J. 2007,
\newblock \apjl, 661, L183

\bibitem[Morrell \& Levato, 1991]{1991ApJS...75..965M}
Morrell, N. \& Levato, H. 1991,
\newblock \apjs, 75, 965

\bibitem[Otero, 2007]{2007OEJV...72....1O}
Otero, S.~A. 2007,
\newblock Open European Journal on Variable Stars, 72, 1

\bibitem[Perryman \& ESA, 1997]{1997hity.book.....P}
Perryman, M. A.~C. \& ESA 1997,
\newblock The {HIPPARCOS} and {TYCHO} {C}atalogues. {A}strometric and
  {P}hotometric {S}tar {C}atalogues {D}erived from the {ESA} {HIPPARCOS}
  {S}pace {A}strometry {M}ission, Vol. 1200 of ESA SP Series (Noordwijk, The
  Netherlands: ESA Pub. Div.)

\bibitem[Pozzo et~al., 2000]{2000MNRAS.313L..23P}
Pozzo, M., Jeffries, R.~D., Naylor, T., Totten, E.~J., Harmer, S., \& Kenyon,
  M. 2000,
\newblock \mnras, 313, L23

\bibitem[Rauw et~al., 2002]{2002AaA...394..993R}
Rauw, G. et~al. 2002,
\newblock \aap, 394, 993

\bibitem[Roberts, 2001]{amos2001}
Roberts, Jr., L.~C. 2001,
\newblock in Proc. 2001 AMOS Technical Conference, ed. P.~Kervin, L.~Bragg, \&
  S.~Ryan (Maui, HI: Maui Econ. Devel. Board),  326

\bibitem[Roberts \& Neyman, 2002]{RobertsNeyman}
Roberts, Jr., L.~C. \& Neyman, C.~R. 2002,
\newblock \pasp, 114, 1260

\bibitem[Roberts et~al., 2007]{2007AJ....133..545R}
Roberts, Jr., L.~C., Turner, N.~H., \& {\SortNoop{Brummelaar}}{ten Brummelaar},
  T.~A. 2007,
\newblock \aj, 133, 545

\bibitem[Robin et~al., 2003]{2003AaA...409..523R}
Robin, A.~C., Reyl{\'e}, C., Derri{\`e}re, S., \& Picaud, S. 2003,
\newblock \aap, 409, 523

\bibitem[Rossiter, 1955]{1955POMic..11....1R}
Rossiter, R.~A. 1955,
\newblock Publications of Michigan Observatory, 11, 1

\bibitem[See, 1898]{1898AJ.....18..181S}
See, T. J.~J. 1898,
\newblock \aj, 18, 181

\bibitem[Seymour et~al., 2002]{2002AJ....123.1023S}
Seymour, D.~M., Mason, B.~D., Hartkopf, W.~I., \& Wycoff, G.~L. 2002,
\newblock \aj, 123, 1023

\bibitem[Silbernagel, 1931]{1931AN....241...33S}
Silbernagel, E. 1931,
\newblock Astronomische Nachrichten, 241, 33

\bibitem[Struve, 1837]{1837AN.....14..249S}
Struve, F. G.~W. 1837,
\newblock Astronomische Nachrichten, 14, 249

\bibitem[Wallenquist, 1934]{Wallenquist1934}
Wallenquist, A. 1934,
\newblock Ann. Bosscha Obs. Lembang, 6,
\newblock pt. 2

\bibitem[Wegner, 1994]{1994MNRAS.270..229W}
Wegner, W. 1994,
\newblock \mnras, 270, 229

\bibitem[{\SortNoop{Wit}}{de Wit} et~al., 2004]{2004AaA...425..937D}
{\SortNoop{Wit}}{de Wit}, W.~J., Testi, L., Palla, F., Vanzi, L., \& Zinnecker,
  H. 2004,
\newblock \aap, 425, 937

\bibitem[{\SortNoop{Wit}}{de Wit} et~al., 2005]{2005AaA...437..247D}
{\SortNoop{Wit}}{de Wit}, W.~J., Testi, L., Palla, F., \& Zinnecker, H. 2005,
\newblock \aap, 437, 247

\bibitem[Worley, 1990]{1990ebua.conf..419W}
Worley, C.~E. 1990,
\newblock in Errors, Bias and Uncertainties in Astronomy, ed. C.~Jaschek \&
  F.~Murtagh (Cambridge: Cambridge Univ. Press),  419

\bibitem[Wycoff et~al., 2006]{2006AJ....132...50W}
Wycoff, G.~L., Mason, B.~D., \& Urban, S.~E. 2006,
\newblock \aj, 132, 50

\bibitem[Zinnecker \& Yorke, 2007]{2007ARAaA..45..481Z}
Zinnecker, H. \& Yorke, H.~W. 2007,
\newblock \araa, 45, 481

\end{thebibliography}
\end{document}